\documentclass[prb,aps,twocolumn,superscriptaddress,eqsecnum,amsmath,amssymb,showpacs,floatfix]{revtex4-1}
\usepackage{graphicx}
\usepackage{dcolumn}
\usepackage{bm}
\usepackage{placeins}

\begin{document}

\title{Defect states and excitations in a Mott insulator with orbital degrees of freedom:
       Mott-Hubbard gap versus optical and transport gaps in doped systems}

\author{     Adolfo Avella}
\affiliation{ Max-Planck-Institut f\"ur Festk\"orperforschung,
              Heisenbergstrasse 1, D-70569 Stuttgart, Germany }
\affiliation{Dipartimento di Fisica ``E.R. Caianiello'', 
             Universit\`{a} degli Studi di Salerno, I-84084 Fisciano (SA), Italy}
\affiliation{CNR-SPIN, UoS di Salerno, I-84084 Fisciano (SA), Italy}
\affiliation{Unit\`{a} CNISM di Salerno, Universit\`{a} degli Studi di Salerno, 
             I-84084 Fisciano (SA), Italy}

\author{Peter Horsch}
\affiliation{ Max-Planck-Institut f\"ur Festk\"orperforschung,
              Heisenbergstrasse 1, D-70569 Stuttgart, Germany }

\author {     Andrzej M. Ole\'s}
\affiliation{ Max-Planck-Institut f\"ur Festk\"orperforschung,
              Heisenbergstrasse 1, D-70569 Stuttgart, Germany }
\affiliation{ Marian Smoluchowski Institute of Physics, Jagellonian
              University, Reymonta 4, PL-30059 Krak\'ow, Poland }

\date{31 January 2013}

\begin{abstract}
We address the role played by charged defects in doped Mott insulators 
with active orbital degrees of freedom. It is observed that defects 
feature a rather complex and rich physics, which is well captured by a 
degenerate Hubbard model extended by terms that describe crystal-field 
splittings and orbital-lattice coupling, as well as by terms generated 
by defects such as the Coulomb potential terms that act both on doped 
holes and on electrons within occupied orbitals at undoped sites.
We show that the multiplet structure of the excited states generated in 
such systems by strong electron interactions is well described within 
the unrestricted Hartree-Fock approximation, once the symmetry breaking 
caused by the onset of magnetic and orbital order is taken into account.
Furthermore, we uncover new spectral features that arise within the 
Mott-Hubbard gap and in the multiplet spectrum at high energies due to 
the presence of defect states and strong correlations. 
These features reflect the action on 
electrons/holes of the generalized defect potential that affects charge 
and orbital degrees of freedom, and indirectly also spin ones.
The present study elucidates the mechanism behind the Coulomb gap 
appearing in the band of defect states and investigates the dependence 
on the electron-electron interactions and the screening by the orbital 
polarization field.
As an illustrative example of our general approach, we present explicit 
calculations for the model describing three $t_{2g}$ orbital flavors in 
the perovskite vanadates doped by divalent Sr or Ca ions, such as in
La$_{1-x}$Sr$_x$VO$_3$ and Y$_{1-x}$Ca$_x$VO$_3$ systems. We analyze 
the orbital densities at vanadium ions in the vicinity of defects, 
and the excited defect states which determine the optical and transport 
gaps in doped systems.
\\
{\it Published in: Physical Review B} {\bf 87}, 045132 (2013).

\end{abstract}


\maketitle

\section{Introduction\label{sec:intro}}

Doped vanadium perovskites show a very rich behavior that has been
mainly explored by experimentalists during the last decade. 
\cite{Kas93,Miy00,Fuj05,Fuj08,Sag08,Fuj10,Uch11,Yan11} In contrast to 
the high-T$_c$ cuprates,\cite{Dam03,Lee06,Ave07,Oga08,Phi10,Sca12} the 
perovskite vanadates represent quite generally systems with an active 
orbital degree of freedom.\cite{Ima98,Nag00,Ole05,Hot06} The particular 
excitement with the $R$VO$_3$ perovskite mother compounds, where 
$R$=Lu,Yb,$\cdots$,La, is the quantum interplay of spin-orbital degrees 
of freedom,
\cite{Kha01,Sir03,Hor03,Kha04,Kha05,Ole06,Ole07,Hor08,Sir08,Ole12} 
which is only under partial control of weak Jahn-Teller (JT) and other 
couplings.\cite{Hwa12} The moderate coupling to the lattice 
results from the $t_{2g}$ character of the outer valence electrons,
\cite{Zaa93} and explains that the orbital quantum 
dynamics is not quenched --- a property vanadates share with other 
transition metal oxides like for example titanates\cite{Kha00,Jac08}
and iridates.\cite{Jac09,Kim12}
The phase diagrams of doped vanadates\cite{Fuj05} deduced from the 
resistivity, specific heat and magnetic measurements show a systematic 
decrease of magnetic interactions and the onset of the metallic 
behavior only at rather high doping,\cite{Fuj05} in contrast, for 
instance, to the high-$T_c$ cuprates where the metal-insulator 
transition (MIT) is found at a few percent of hole doping.\cite{Che91} 
In La$_{1-x}$Sr$_x$VO$_3$, perhaps the most investigated doped vanadium 
oxide, the MIT is at $x\simeq 0.18$ Sr doping,\cite{Miy00} while 
Y$_{1-x}$Ca$_x$VO$_3$ is insulating up to $x\simeq 0.50$.\cite{Fuj08} 
This latter compound undergoes a transition from the $GC$ phase to the 
$CG$ phase\cite{notecg} at rather low doping $x\sim 0.02$ within the 
insulating phase. It was pointed out\cite{Kha05} that the double 
exchange\cite{deG60,vdB99,Hor99,Ber12} is the mechanism responsible for 
this transition in antiferromagnetic (AF) phases, similar to the 
transition from the $G$-type AF ($G$-AF) to $C$-type AF ($C$-AF) order 
in electron doped manganites.\cite{Ole11} In addition, the change of 
superexchange interaction in the vicinity of the holes\cite{notej} 
bound to the defects has a strong influence on the relative energy of 
these states.\cite{Hor11}
Remarkably the $CG$ spin-orbital order persists up to the MIT.

The evolution of optical spectra with doping for these two 
vanadium perovskites,\cite{Fuj08} La$_{1-x}$Sr$_x$VO$_3$ and 
Y$_{1-x}$Ca$_x$VO$_3$, shows that the defects lead to impurity 
states which are responsible for an absorption band deep inside the
Mott-Hubbard (MH) gap.\cite{Hor11} This suggests that 
bound small polarons are the cause of the MIT even at high 
doping concentrations.\cite{Fuj08} It is eventually the growth of the 
mid-infrared absorption under increasing doping and the gradual shift 
of this absorption toward zero energy, which accompanies the MIT. 
These experimental findings call for a theoretical approach that would 
address the nature of defects and the changes in the magnetic and 
orbital structure induced by them in a Mott 
insulator. Doping in combination with strong correlations leads to 
spectral weight transfer,\cite{Hor11} 
while on the other hand the size of the MH gap is not
significantly influenced by doping.\cite{Fuj08} We discuss these issues 
in the present paper and point out that the defect states here should be 
distinguished from those observed within the MH gap in other 
circumstances such as the excitonic states in LiCuVO$_4$,\cite{Mat09} 
or the states generated simply by doping an orbitally ordered Mott 
insulator in absence of defects.\cite{Wro12}

The description of defect states in systems with strongly correlated
electrons is an outstanding problem in the theory of condensed matter,
\cite{Gan05} and has been addressed mainly within the Hartree-Fock (HF) 
approach,\cite{Miz00,Eva03,Che09,Hor11} and the methods based on density 
functional theory.\cite{Car06,Kon09,Pav12}
It involves for example questions like: 
(i) What is the nature of defect states in a strongly correlated system, 
i.e., as compared to defects in usual semiconductors or insulators?
\cite{Yu96,Que98,Dra07}
(ii) What happens to the MH gap in the presence of defects? 
(iii) Which are the new features in orbital degenerate MH insulators,
as for example in the $t_{2g}$ transition metal oxides with active 
orbital degree of freedom, in contrast to doped high-T$_c$ 
superconductors?
(iv) Which are practicable methods to perform reliable calculations for 
MH insulators that can be extended to take into account defects and 
disorder?
The answer to all these questions is a formidable challenge as the 
defects may strongly affect the subtle interplay of several essential 
degrees of freedom, namely  spin, orbital, lattice and charge,
as we shall see below.

It is remarkable that the theoretical analysis of doping in transition
metal oxides, including manganites and also superconducting cuprates, 
is usually performed in the metallic regime where the effect of doping 
on the properties of a material with strongly correlated electrons 
leads solely to the modification of the electronic filling in party 
filled $3d$ orbitals.\cite{Ima98} This simplification is not possible 
in the insulating regime where local defect fields are not screened, 
leading to defect states, as for example in the case of Ca defects in 
the lattice of Y ions in Y$_{1-x}$Ca$_x$VO$_3$.\cite{Hor11} Here, we 
explore further this idea and analyze a model with three $t_{2g}$ 
orbital flavors. This is a generalization of the frequently used 
two-flavor model (with active $\{xz,yz\}$ orbitals) for $R$VO$_3$ 
compounds,\cite{Kha01,Sir03,Hor03,Kha04,Kha05,Ole07,Hor08,Hor11} 
motivated by earlier electronic structure studies.\cite{Saw96,Sol06,And07} 
The first electron at each V$^{3+}$ ion occupies the $xy$ orbital, with 
energy lowered by the crystal field, while the second one occupies one 
of two degenerate orbitals, either the $yz$ or $xz$, resulting in an 
$xy^1(yz/zx)^1$ local configuration.

The purpose of this paper is to introduce a systematic unbiased approach
that allows one to study defect states in strongly correlated materials. 
As we show below, the interactions, which arise by introducing charged
defects into a Mott insulator, are numerous and all of them are 
necessary in a realistic approach that is capable of predicting 
experimental behavior of a doped material. We adopt here a multiband 
Hubbard model\cite{Ole83} 
that is supplemented by several terms responsible for the influence of
the lattice and defect states on the electronic structure. We use it
below for the description of $t_{2g}$ electrons and generalize thereby 
earlier approaches.\cite{Ole05,Bou09,Dag10} The electronic structure 
will be analyzed using the unrestricted HF approximation,
\cite{Pratt,Miz96,Wen10} which is known to be reliable in systems with
spontaneously broken symmetry, for instance for states with magnetic 
order in multiband models for manganites,\cite{Miz00} CuO$_2$ planes
of high temperature superconductors,\cite{Ole89} and low-dimensional 
cuprates,\cite{Grz91} and may serve to describe complex types of order 
such as stripe or spiral structures.\cite{Kim09,Sei11,Sol11}
As we demonstrate below, charged defects are the source of impurity 
states which appear as new features  within the MH gap, while the MH gap
itself is essentially unaffected in the low doping regime.\cite{Fuj08}
Hence the defect states are responsible for the  in-gap absorption
in the optical conductivity at low doping.
At the same time, we demonstrate that the HF method used here is well
designed to treat simultaneously phenomena that arise at distinct
energy scales, the high energy $\sim 1$ eV related to the (on-site and
intersite) Coulomb interaction around the defect states and the low
energy scale $\sim 0.1$ eV that is related to the orbital physics and
controls electronic transport in doped materials.

The problem addressed here pertains to general transition metal oxides 
with perovskite or layered structure, which feature
a variety of interesting phenomena due to strong electron correlations.
They include MITs in undoped systems, several phenomena with long-range 
(LR) coherence in doped systems, such as high-$T_c$ superconductivity,
\cite{Lee06,Ave07,Oga08,Phi10,Sca12} colossal
magnetoresistance, phases with magnetic and orbital order, and so on.
In particular, the doped manganese oxides with colossal 
magnetoresistance show several magnetic phase transitions
\cite{Dag01,Wei04,Tok06} that have attracted a broad interest
in the interplay between spin, orbital and charge degrees of freedom
in strongly correlated electron systems. In this class of systems, 
orbitals couple strongly to the lattice and this coupling supports the 
superexchange,\cite{Fei99} leading to well separated energy scales for 
the structural and magnetic transition.\cite{Tok06} In contrast,
the $R$VO$_3$ perovskites are a
challenge for the theory of spin-orbital systems as the magnetic and
orbital order occur here in the same range of temperatures, or even
simultaneously as in LaVO$_3$.\cite{Miy03,Fuj10}

A common feature of the $R$VO$_3$ perovskites is the onset of the
$G$-type alternating orbital ($G$-AO) order below the characteristic
orbital ordering temperature $T_{\rm OO}$, which is in these compounds
relatively low, $T_{\rm OO}\simeq 200$ K, and exhibits non-monotonous
dependence on the ionic radius of $R$ ions.\cite{Miy03,Fuj10} The
temperature $T_{\rm OO}$ comes close to the N\'eel temperature
$100<T_{\rm N1}<140$ K for the magnetic transition to the $C$-type 
antiferromagnetic ($C$-AF) phase. This phase competes with another AF 
phase in the $R$VO$_3$ systems with small radii of $R$ ions.\cite{Fuj10} 
In such cases, this latter $GC$ phase is stable at low temperature,
while the former $CG$ phase takes over when temperature increases. 
The best known example of this behavior is YVO$_3$ with a transition
temperature $T_{\rm N2}=77$ K. It has been shown in Ref. 
\onlinecite{Hor03} that relativistic spin-orbit interaction is important 
to describe the  reorientation of spins
associated with the two different types of order at this 1st order
phase transition. A remarkable reversal of the magnetization direction
\cite{Ren00,Nog00} inside the CG-phase with increasing temperature
could not be understood in the theory until now.

The principal difficulty in the theory of the $R$VO$_3$ perovskites is
the quantum interplay of spin-orbital degrees of freedom,\cite{Kha01} 
which manifests itself in leading contributions that emerge due to 
spin-orbital entanglement.\cite{Ole06,Ole12} Quantum effects 
associated with joint spin-orbital dynamics play a role at zero and 
at finite temperatures\cite{Miy02,Ulr03,Tun08} and the consequences 
have been observed in several experiments:\cite{Ole12}
(i) the temperature dependence of the optical spectral weights,
\cite{Miy02}
(ii) the apparent breakdown of Goodenough-Kanamori rules for magnetic 
interactions in the $CG$ phase [unlike in other orbitally degenerate 
systems, ferromagnetic (FM) interactions are 
here stronger than the AF ones],\cite{Ulr03}
(iii) the orbital-Peierls dimerization of FM interactions in the $C$-AF 
phase\cite{Sir03,Hor03,Sir08} observed in the neutron scattering 
in YVO$_3$,\cite{Ulr03} and also in LaVO$_3$,\cite{Tun08} as well as
(iv) the interplay between the spin and orbital correlations, which lead
to the phase diagram of the $R$VO$_3$ perovskites.\cite{Fuj10}
The Peierls instability of FM
chains towards dimerization occurs when spin and orbital degrees of
freedom are entangled at finite temperature,\cite{Sir08} and
a previous theoretical study showed that charged defects favor
orbital-Peierls dimerization in the small doping regime.\cite{Hor11}
Furthermore orbital
fluctuations and their competition with orbital-lattice coupling play a
crucial role for the explanation of the observed non-monotonous
dependence of the orbital transition temperature on the radius $r_R$ of
$R$ ions along the $R$VO$_3$ series.\cite{Hor08}
Other studies support the picture of strong orbital fluctuations
and their decisive impact on the observed physical properties.
\cite{Ree06,Goo07,Maz08,Zho09}
We show below that in spite of the above quantum nature of the 
spin-orbital order in the $R$VO$_3$ perovskites, the HF approach is
well designed to capture the essential features of charge defects.

The paper is organized as follows. First, in Sec. \ref{sec:3band},
we introduce a degenerate Hubbard model for $t_{2g}$ electrons in
the $R$VO$_3$ perovskites and generalize it to the doped materials,
such as Y$_{1-x}$Ca$_x$VO$_3$ with Ca$^{2+}$ charge defects replacing 
some Y$^{3+}$ ions. The model includes local and intersite Coulomb 
interactions, and the Coulomb potentials induced by Ca defect states, 
which change locally the electronic structure of $t_{2g}$ electrons 
within V($3d$) orbitals. We describe the treatment of this model in 
the unrestricted HF approximation in Sec. \ref{sec:HF}. In Sec. 
\ref{sec:MIT}, we present the electronic structure obtained for the 
undoped $CG$ and $GC$ phases in Mott insulators for the typical 
parameters of $R$VO$_3$. As explained there, the electronic structures 
provide a realistic description and include the multiplet structure of 
$d^3$ excited states. The optimized $t_{2g}$ orbitals, at a vanadium ion 
being the nearest neighbor of a Ca defect, change to a new orthogonal 
basis due to the defect-orbital polarization interaction, as shown in 
Sec. \ref{sec:atom}. Next, we analyze the electron distribution on 
a bond next to a Ca defect in the $CG$ and $GC$ phases in Sec. 
\ref{sec:bondd}, and compare the unrestricted HF approach with the exact 
diagonalization (ED). We also analyze one-particle local charge 
excitations in Sec. \ref{sec:exci} and demonstrate that their energies 
are well reproduced around the defect states when the 
intratomic Coulomb interaction $U$ increases, see Sec. \ref{sec:bondu}. 
Next, we consider the electronic structure for the defect states
in the $CG$ phase in the dilute doping limit in Sec. \ref{sec:dopd},
and introduce in Sec. \ref{sec:land} the 
\emph{order parameter landscape} in a doped system.
We study the effects of a finite orbital polarization
in Sec. \ref{sec:three} and investigate the changes induced by longer 
range Coulomb interactions in Sec. \ref{sec:dopv}. 

This analysis is followed by the discussion of the
electronic structure in a correlated material with defect states,
where we address the gaps observed in the optical spectroscopy in Sec.
\ref{sec:gap} and the dependence of the electron density in $xy$ 
orbitals on the defect concentration in Sec. \ref{sec:nc}. Finally, the 
paper is summarized in Sec. \ref{sec:summa}, where we also present an 
outlook at possible future applications of the general unrestricted HF 
method introduced in the paper in doped transition metal oxides. 
The Appendix highlights the importance of Fock terms in the present
problem and addresses the concept of optimized orbitals in the vicinity 
of a charge defect.

\section{Mott insulator with defect states}
\label{sec:model}

\subsection{Degenerate Hubbard model for $t_{2g}$ electrons}
\label{sec:3band}

We begin with introducing the multiband Hubbard model for $t_{2g}$ 
electrons designed to describe doped $R$VO$_{3}$ perovskites, such as 
the Y$_{1-x}$Ca$_{x}$VO$_{3}$ compounds. This effective model includes 
only $t_{2g}$ orbitals at vanadium ions V$^{3+}$, coupled by effective 
$d-d$ hopping elements along V--O--V bonds. Below, we consider how this 
picture is modified in presence of charged defects and, in particular, 
when not only the electron density within the $t_{2g}$ orbitals of 
vanadium ions changes, but also the presence of Ca defects introduces 
local interactions acting on $t_{2g}$ electrons at vanadium ions in 
their vicinity. 

The multiband Hamiltonian that describes the MH physics of
quasi-degenerate $t_{2g}$ electrons and the perturbations by the defect 
potentials consists quite generally of one-electron terms ${\cal H}_{0}$ 
and two-electron (Coulomb and JT) interactions 
${\cal H}_{{\rm int}}$:
\begin{eqnarray}
{\cal H} & = & {\cal H}_{0}+{\cal H}_{{\rm int}},\label{model}\\
{\cal H}_{0} & = & H_{t}+H_{{\rm CF}}+H_{{\rm def}},\label{H0}\\
{\cal H}_{{\rm int}} & = & H_{U}+H_{V}+H_{{\rm JT}}.\label{Hint}
\end{eqnarray}
The one-electron part (\ref{H0}) is composed of the following  terms: 
(i) the kinetic energy ($H_{t}$),
(ii) crystal-field (CF) splitting ($H_{{\rm CF}}$), and
(iii) the perturbations generated by defects ($H_{{\rm def}}$). 
Here, $H_{{\rm def}}$ includes two terms: 
\begin{equation}
{\cal H}_{\rm def}  =  H_{{\rm imp}}+H_{{\rm pol}},\label{Hdef}
\end{equation}
the Coulomb potentials of the charged impurities  $H_{{\rm imp}}$
and the orbital polarization term $H_{{\rm pol}}$. These two terms 
arise due to the presence of Ca defects, and they constitute effective 
fields that act on the vanadium ions in the neighborhood of defects. 
They have been introduced in a simplified (two-flavor) model with two 
active $t_{2g}$ orbital flavors $\{yz,zx\}$,\cite{Hor11} see below.

The electron interactions ${\cal H}_{{\rm int}}$ (\ref{Hint}) are given 
by: (i) local Coulomb interactions ($H_{U}$), 
(ii) intersite Coulomb interactions ($H_{V}$), and 
(iii) the JT effective interactions between orbitals which are 
induced by lattice distortions ($H_{{\rm JT}}$).
In the case of $R$VO$_{3}$ perovskites considered here, the local Coulomb 
interactions are quite strong and lead to a Mott insulator with high spin 
($S=1$) and orbital degrees of freedom.\cite{Kha01,Kha04}
Depending on the parameters, the interactions in ${\cal H}_{{\rm int}}$
support a particular type of symmetry-broken phase,\cite{Ole07} and one 
expects coexisting spin and orbital order, either in the form of the $CG$ 
or $GC$ phase, in agreement with experimental observations.\cite{Fuj10}

The kinetic energy describes the hopping processes between nearest
neighbor V$^{3+}$ sites on bonds $\left\langle ij\right\rangle $
oriented along one of three cubic directions
$\gamma=a,b,c$ in the perovskite lattice,
\begin{equation}
H_{t}=\sum_{\left\langle ij\right\rangle \parallel\gamma}
\sum_{\alpha,\beta,\sigma} t^{\gamma}_{\alpha\beta}
\left(c_{i\alpha\sigma}^{\dag}c_{j\beta\sigma}^{}
+c_{j\beta\sigma}^{\dag}c_{i\alpha\sigma}^{}\right).\label{ht}
\end{equation}
Here $c_{i\alpha\sigma}^{\dag}$ is the electron creation operator at the 
V$^{3+}$ ion at site $i$, with orbital flavor $\alpha$ and spin 
$\sigma=\uparrow,\downarrow$. The summation runs over all bonds
$\langle ij\rangle{\parallel}\gamma$. The effective hopping $t$ 
originates from two subsequent $d-p$ hopping processes via the 
intermediate O$(2p_{\pi})$ orbital along each V--O--V bond. It follows 
from the charge-transfer model with $p-d$ hybridization $t_{pd}$ and 
charge-transfer energy $\Delta$,\cite{Zaa93}
that $t$ is finite only between two identical $t_{2g}$ orbitals,
labeled $\alpha(\gamma)$, that are active along a given bond
$\langle ij\rangle{\parallel}\gamma$. The third orbital, which lies in 
the plane perpendicular to the $\gamma$
axis, is inactive as the hopping processes vanish here by symmetry.
This motivates the convenient notation used hereafter,\cite{Kha01}
\begin{equation}
|a\rangle\equiv|yz\rangle,\hskip.7cm|b\rangle\equiv|xz\rangle,
\hskip.7cm|c\rangle\equiv|xy\rangle,\label{abc}
\end{equation}
where the orbital $|\gamma\rangle$ inactive along a cubic direction 
$\gamma$ is labeled by its index. Thus, if we consider an idealized case 
without lattice distortions and without defects, the hopping conserves 
the $t_{2g}$ orbital flavor, 
\begin{equation}
t^{\gamma}_{\alpha\beta}= 
-t \delta_{\alpha\beta} (1-\delta_{\gamma\alpha}).
\end{equation}
Using the charge-transfer model\cite{Zaa93} one estimates the hopping 
element $t=t_{pd}^{2}/\Delta\simeq0.2$ eV --- this value appears to be 
consistent with that deduced from the electronic structure calculations.
\cite{And07}

The $t_{2g}$ orbital states are non-equivalent in the $R$VO$_{3}$
perovskites due to the GdFeO$_{3}$-like distortions,\cite{Miz99} and one 
finds that the $c$ orbitals are occupied at every site. This is a
consequence of the CF splitting term, see Eq. (\ref{model}), which 
favors $c$ orbital occupation,
\begin{equation}
H_{{\rm CF}}=-\varepsilon_{c}^{0}\sum_{i}n_{ic}\,,
\label{hcf}
\end{equation}
where $n_{i\alpha}=\sum_{\sigma}n_{i\alpha\sigma}$, with
$n_{i\alpha\sigma}=c_{i\alpha\sigma}^{\dag}c_{i\alpha\sigma}^{}$,
is the electron density operator for spin-orbital $\{\alpha\sigma\}$ at 
site $i$. We estimated,\cite{Hor08} $\varepsilon_{c}^{0}\sim t$ --- 
hence the $c$ orbital is filled by one electron
and the second electron at any V$^{3+}$ ion occupies one of the orbitals
in the $\{a,b\}$ doublet, leading to a $c_{i}^{1}(a,b)_{i}^{1}$
configuration at each site $i$. This broken symmetry state, which
corresponds to electron densities
\begin{equation}
\langle n_{ic}\rangle=1,\hskip.7cm\langle n_{ia}+n_{ib}\rangle=1,\label{frozen}
\end{equation}
within $t_{2g}$ orbitals at undoped V$^{3+}$ ions, is stable in a Mott 
insulator even for values of $\varepsilon_{c}$ as small as $0.5t$ and 
justifies the two-flavor model.\cite{Hor11} Note, however, that one 
expects that the cubic symmetry with $n_{i\gamma}=2/3$ is restored at 
high temperatures,\cite{Kha01,Ole07} but this situation will not be 
analyzed here.

The defect terms in $H_{{\rm 0}}$, Eq. (\ref{H0}), describe 
Coulomb interactions of the charged defects and the $t_{2g}$ electrons 
within orbitals at neighboring vanadium
ions and act on them as effective fields. As explained elsewhere,
\cite{Hor11} when an Y$^{3+}$ ion in YVO$_{3}$ is replaced by a
Ca$^{2+}$ impurity in the doped compound, Y$_{1-x}$Ca$_{x}$VO$_{3}$,
the lattice is locally disturbed and the impurity acts as an effective
\textit{negative charge\/} in the sublattice of Y ions --- it is located 
in the center of the V$_{8}$ cube surrounding the defect site, see Fig.
1 of Ref. \onlinecite{Hor11}. Such a defect is therefore responsible
for additional Coulomb interaction terms between the impurity charge
and the electron charges in $t_{2g}$ orbitals of surrounding vanadium
ions. The first additional term is the Coulomb potential due to the
Ca impurity, $H_{{\rm imp}}$, which describes the Coulomb interaction
between the defect at site $m$ with effective negative charge $Q_{m}=e$
(here we adopt $e=1$) and the total $t_{2g}$ electron charge density
operator at the V ion at site $i$,
\begin{equation}
H_{{\rm imp}}=\sum_{\mathbf{m}\in{\cal D}}\sum_{i}
W\left((|\mathbf{r}_{i}-\mathbf{R}_{m}|\right)\, n_{i}\,,
\label{Himp}
\end{equation}
where $n_{i}=\sum_{\alpha}n_{i\alpha}$ is the electron density operator.
Here the first sum runs over the Ca defects labeled with $m\in{\cal D}$,
where the set ${\cal D}$ contains all lattice defect sites at the
considered doping $x$ in Y$_{1-x}$Ca$_{x}$VO$_{3}$. The second
index $i$ labels V ions at distance $|\mathbf{r}_{i}-\mathbf{R}_{m}|$
from the considered defect site $m$. In general, the coordinates
of the defects $\{\mathbf{R}_{m}\}$ are statistically distributed,
although defects will also feel some repulsion that avoids their 
clustering. 
The defect potential of an impurity with negative charge $Q_D=e$
is of Coulomb type $W(r)=eQ_D/\epsilon_c r$, where $\epsilon_c$
is the dielectric constant (of core electrons).\cite{Hor11}
Below we shall employ a truncated defect potential
\begin{equation}
W\left(|\mathbf{r}_{i}-\mathbf{R}_{m}|\right)=
\begin{cases}
&V_{D},\; \textrm{if}\; i\in\mathcal{C}^1_{m},\label{VD}\\
&0,\;\;\; \textrm{otherwise}
\end{cases}
\end{equation}
i.e., $i$ should belong to the V$_8$ cube 
${\cal C}^1_m\equiv{\cal C}^1({\bf R}_m)$ formed by eight 
nearest-neighbor V sites surrounding the Ca defect at site $m$ (V$_{8}$ 
cube). With a dielectric constant $\epsilon_c\simeq 5$ for YVO$_3$
one finds for the  defect potential $V_D\simeq  1$ eV.\cite{Hor11}
We note that the  truncated potential is introduced here  
as it provides lateron a more transparent interpretation of the spectra
and facilitates the analysis of the energy shifts of defect states.

\begin{figure}[t!]
\includegraphics[width=8cm]{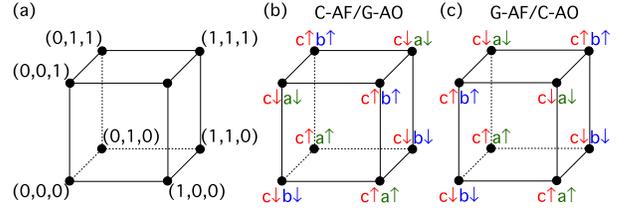}
\caption{(Color online) 
(a) Reference coordinates $(klm)$ of vanadium ions in a representative 
V$_{8}$ cube in YVO$_{3}$. 
Phases with spin-orbital order of perovskite vanadates considered in 
this paper have the electron occupancies in the two 
magnetic phases shown schematically in: 
(b) $C$-AF/$G$-AO phase; 
(c) $G$-AF/$C$-AO phase. 
The occupied orbitals $\{a,b,c\}$ and their spin components 
$\{\uparrow,\downarrow\}$ are indicated at each corner of the cube.}
\label{fig:1}
\end{figure}

The last term in Eq. (\ref{H0}) is the orbital polarization term,
$H_{{\rm pol}}$. The V ions surround a defect site and form a V$_{8}$ 
cube,\cite{Hor11} occupying the positions shown in Fig. \ref{fig:1}(a). 
$H_{{\rm pol}}$ originates from the quadrupolar component of 
electrostatic field generated by a charge defect within the lattice of 
V$^{3+}$ ions. At each vanadium ion 
the Coulomb repulsion between the defect charge and $t_{2g}$ electrons
favors the occupation of the linear combinations of $t_{2g}$ orbitals
that maximize the average distance of the electronic charge from the
defect.

Consequently, the defect-orbital interaction, called below orbital
polarization, takes the form:
\begin{equation}
H_{{\rm pol}}=D\sum_{\mathbf{m}\in{\cal D}}
\sum_{{i\in\mathcal{C}_{m}\atop \alpha\neq\beta}}
\lambda_{\alpha\beta}(\mathbf{r}_{i}-\mathbf{R}_{m})\left(c_{i\alpha\sigma}^{\dag}
c_{i\beta\sigma}^{}+c_{i\beta\sigma}^{\dag}c_{i\alpha\sigma}^{}\right).
\label{HD}
\end{equation}
The coefficients $\{\lambda_{\alpha\beta}(\mathbf{r}_{i}-\mathbf{R}_{m})\}=\pm1$
depend on the pair of considered orbitals $\alpha\beta$ and on the
direction $\mathbf{r}_{i}-\mathbf{R}_{m}$; they are selected to minimize
the Coulomb repulsion. This is achieved by
\begin{eqnarray}
\lambda_{ab}\left(\mathbf{r}_{i}-\mathbf{R}_{m}\right) & = & \left\{ 
\begin{array}{ccc}
1 & \text{if} & \mathbf{r}_{i}-\mathbf{R}_{m}\parallel(111),(11\bar{1}),\\
-1 & \text{if} & \mathbf{r}_{i}-\mathbf{R}_{m}\parallel(\bar{1}11),(1\bar{1}1),
\end{array}\right.\nonumber \\
\\
\lambda_{ac}\left(\mathbf{r}_{i}-\mathbf{R}_{m}\right) & = & \left\{ 
\begin{array}{ccc}
1 & \text{if} & \mathbf{r}_{i}-\mathbf{R}_{m}\parallel(111),(\bar{1}11),\\
-1 & \text{if} & \mathbf{r}_{i}-\mathbf{R}_{m}\parallel(1\bar{1}1),(11\bar{1}),
\end{array}\right.\nonumber \\
\\
\lambda_{bc}\left(\mathbf{r}_{i}-\mathbf{R}_{m}\right) & = & \left\{ 
\begin{array}{ccc}
1 & \text{if} & \mathbf{r}_{i}-\mathbf{R}_{m}\parallel(111),(1\bar{1}1),\\
-1 & \text{if} & \mathbf{r}_{i}-\mathbf{R}_{m}\parallel(\bar{1}11),(11\bar{1}).
\end{array}\right.\nonumber \\
\end{eqnarray}
Note that each direction along one of the diagonals of the cube involves
two vanadium ions.

The local Coulomb interactions between $t_{2g}$ electrons at V$^{3+}$
ions are described by,\cite{Ole83}
\begin{eqnarray}
H_{U} & = & U\sum_{i,\alpha}n_{i\alpha\uparrow}n_{i\alpha\downarrow}
+  \left(U-\frac{5}{2}J_{H}\right)\sum_{i,\alpha<\beta}n_{i\alpha}n_{i\beta}
\nonumber \\
& - & 2J_{H}\sum_{i,\alpha<\beta}\mathbf{S}_{i\alpha}\cdot\mathbf{S}_{i\beta}
+ J_{H}\sum_{i,\alpha\neq\beta}c_{i\alpha\uparrow}^{\dag}
c_{i\alpha\downarrow}^{\dag}c_{i\beta\downarrow}c_{i\beta\uparrow}.
\nonumber \\
\label{HU}
\end{eqnarray}
Here the spin operators for orbital $\alpha$ at site $i$, 
$\mathbf{S}_{i\alpha}$,
are defined through the Pauli matrices, $\{\sigma_x,\sigma_y,\sigma_z\}$, 
and are related to fermion operators in the standard way, i.e.,
\begin{equation}
S_{i\alpha}^{+}\equiv c_{i\alpha\uparrow}^{\dagger}c_{i\alpha\downarrow}
\,,\hskip.5cmS_{i\alpha}^{z}\equiv\frac{1}{2}(n_{i\alpha\uparrow}
-n_{i\alpha\downarrow})\,.
\label{S+z}
\end{equation}
The first term in Eq.~(\ref{HU}) describes the intraorbital Coulomb
interaction $U$ between electrons with antiparallel spins; the second
one stands for Hund's exchange $J_{H}$. These two parameters $\{U,J_{H}\}$
suffice to describe local Coulomb interactions between $t_{2g}$ electrons
considered here and are determined by the effective Racah parameters
$\{A,B,C\}$;\cite{Gri71} the remaining terms are the interorbital
Coulomb density-density interaction, and the so-called ``pair-hopping''
term which acts in the subspace of doubly occupied orbitals. The choice
of coefficients in Eq. (\ref{HU}) guarantees that the interactions
obey the rotational invariance in the orbital space.\cite{Ole83}
Note that this Hamiltonian is exact when it describes only one 
representation of the cubic symmetry group (either $t_{2g}$ or $e_{g}$) 
for $3d$ orbitals --- this applies to the present case of  partly 
occupied $t_{2g}$ orbitals in the $R$VO$_{3}$ perovskites. 
In general, however, the interorbital Coulomb and Hund's exchange 
interactions are both anisotropic\cite{Ole05} and preserving rotational 
invariance requires also including on-site three-orbital and 
four-orbital terms;
more details may be found in Refs. \onlinecite{Gri71,Bar00,Bar04,Mcd07}.

The LR Coulomb interaction has the usual expression,
\begin{equation}
H_{V}=\frac{1}{2}\sum_{i\neq j,\alpha\beta}V_{ij}n_{i\alpha}n_{j\beta},
\label{Vee}
\end{equation}
where the Coulomb interaction $V_{ij}$ is parametrized by the nearest
neighbor matrix element $V_{{1}}$ 
\begin{equation}
V_{ij}=\frac{V_{1}}{\left|\mathbf{r}_{i}-\mathbf{r}_{j}\right|}\,,
\label{Vij}
\end{equation}
with coordinates $\mathbf{r}_{i}=(i_a,i_b,i_c)$ and 
$\mathbf{r}_{j}=(j_a,j_b,j_c)$, respectively, given in integer 
representation. We assume that the interaction parameter $V_1$ accounts 
implicitly for the background dielectric function due to the core 
electrons. For convenience we define the nearest neighbor interaction, 
\begin{equation}
V_1=\kappa V_{{\rm ee}},
\label{V1}
\end{equation}
in terms of the electron-electron interaction strength $V_{\rm ee}$, 
where the parameter $\kappa=\frac{\sqrt{3}}{2}$ represents the ratio 
$d/a$; here $a$ the V-V lattice constant and $d$ is the distance between 
the defect and the nearest V neighbor.
In the present paper, we discuss results obtained either without the
intersite Coulomb interaction ($V_{{\rm ee}}=0$) or with the full
LR interaction.

The JT term ${\cal H}_{{\rm JT}}$ describes intersite 
orbital-orbital interactions that follow from lattice distortions and 
concern the orbital doublet $\{a,b\}$. The distortions are both of the 
JT-type and GdFeO$_{3}$-type and jointly induce orbital 
alternation in the $ab$ planes and favor identical orbitals along the 
$c$ axis; these interactions are included here by the following 
expression,\cite{Kha01,Hor11}
\begin{equation}
H_{{\rm JT}}=V_{ab}\sum_{\left\langle ij\right\rangle \parallel a,b}
\tau_{i}^{z}\tau_{j}^{z}-V_{c}\sum_{\left\langle ij\right\rangle \parallel c}
\tau_{i}^{z}\tau_{j}^{z}\,,\label{HJT}
\end{equation}
where
\begin{equation}
\tau_{i}^{z}\equiv\frac{1}{2}(n_{ia}-n_{ib})\label{tauz}
\end{equation}
is the $z$th component of the orbital pseudospin $\tau=1/2$ along the 
$c$ axis influenced by lattice distortions. The interaction parameters
$V_{ab}>0$ and $V_{c}>0$ influence the subtle balance between the
magnetic phases in the $R$VO$_{3}$ perovskites,\cite{Fuj10} supporting 
the $C$-AO order, as realized for instance in the undoped YVO$_{3}$ at 
low temperatures (in the $GC$ phase). These interactions increase within 
the $R$VO$_{3}$ perovskites towards the compounds with small ionic radii, 
and this dependence plays a crucial role in the detailed understanding
of the phase diagram of these perovskites.\cite{Hor08} 

\subsection{Hartree-Fock approximation}
\label{sec:HF}

The degenerate Hubbard Hamiltonian, introduced in the previous Section
to describe $t_{2g}$ electrons in a system with charged defects, can be 
solved exactly by ED only for very small systems.
We present such calculations for a single atom and for a bond in Secs.
III and IV, respectively. Treatment of larger systems is possible only 
after introducing approximations, either in the strong-coupling regime, 
where charge degrees of freedom can be integrated out,\cite{Ole05,Hor11}
or using HF approximation. We adopt here the latter general approach,
which requires the self-consistent calculation of the full density
matrix. We performed calculations that describe Y$_{1-x}$Ca$_{x}$VO$_{3}$
compound in the regime of low doping using finite clusters with defects 
of size $N\times N\times N$, and implemented periodic boundary conditions
(PBC). We present the results obtained for $N=8$ in Secs. V and VI.

The main aim of this paper is to calculate the electronic structure
of strongly correlated and doped MH insulators, i.e.,
including the defects with their complex structure. This implies a
computational challenge due to the simultaneous appearance of the
strong correlation problem and the
perturbations introduced by defects into the 
electronic structure. Certainly it requires reliable approximations
to deal with strong correlations and disorder simultaneously. 
The HF approach is an efficient scheme which maps the interacting 
electron problem onto the problem of a single particle moving in a 
self-consistently determined field that stems from all other electrons.
Whether the HF approach is indeed such a reliable scheme that fulfils 
the requirements stated above is the central topic of this paper.

Before designing this calculation scheme we would like to emphasize
the following aspects (see also the Appendix):\\
(i) It has been shown\cite{Hor11} that a HF factorization when applied 
to a two-orbital model is capable to describe both the MH gap and the 
defect states emerging from the lower Hubbard band due to the defects.
Even more important, the typical multiplet splitting of the transition 
metal ions is obtained if the factorization is performed with respect to 
an optimal local basis set, i.e.,
with occupation numbers either close to 1 or to 0. That is, the local
spin-orbital states should be either occupied or empty.\\
(ii) This requirement leads to complications in the vicinity of defects.
In particular charge defects lead to a rotation of the occupied local 
states, see Sec. \ref{sec:atom}, due to the Coulomb potential 
of a charged defect.\\
(iii) In principle, a local rotation can be specified that removes
the off-diagonal terms and defines a new optimal local basis, i.e., 
orbitals having again integer occupancies (close to 1 or 0). This 
simple scheme, however, is not successful in presence of kinetic energy 
terms. Here the optimal local or Wannier type orbitals are influenced 
not only by local rotation terms, but by these nonlocal terms as well.\\
(iv) It is important to recognize that the full unrestricted HF
scheme, i.e., including all relevant off-diagonal contributions
from interactions, can be used to determine the MH-split bands
in presence of defects. We will show here that this
also holds true for the three-flavor case. In particular, we will
show by comparison with ED that the approach provides a surprisingly 
good description not only of occupied states, but also of the
unoccupied higher multiplet states.\\
(v) Moreover, the local rotated basis may serve as an intermediate
basis and allows one to understand more easily the new features
emerging due to the defects. For example, due to local rotation of 
$t_{2g}$ orbitals the flavor conservation in hopping processes that
contribute to the kinetic energy is lifted in the rotated basis. That 
is, in the vicinity of defects, orbital flavors mix not only due to 
local orbital rotation (off-diagonal CF terms), but also due to the 
kinetic energy.

Except for a special attention to all relevant off-diagonal Fock terms, 
the unrestricted HF procedure applied here is a standard approach for 
tight binding models of interacting electrons.\cite{Bar00,Bar04,Mcd07} 
In general, one decouples each term describing a two-particle 
interaction and replaces it by all possible one-particle terms coupled 
to certain effective mean fields --- density terms in the Hartree 
approximation (HA), supplemented by off-diagonal Fock terms in the HF 
approximation. Such expressions are augmented by double-counting 
correction terms, which are necessary to avoid double counting of the 
interaction energy. We will adopt just Hartree terms in only two cases: 
(i) when the interaction simulates a potential of the lattice and Fock 
terms would simply be out of scope (as for the JT terms), and  
(ii) when there is no one-particle term in the original Hamiltonian that 
can act as a source potential for the related off-diagonal mean fields 
(e.g. $H_{U}$ with respect to spin degree of freedom). 
Note that the JT terms simulate the potential acting on pairs
of neighboring V$^{3+}$ ions due to lattice distortions and 
discriminating between the orbital flavors in the doublet $\{a,b\}$.

As a matter of fact, Fock terms become active only if their off-diagonal 
mean fields are finite. This can happen only if either a one-particle 
term in the original Hamiltonian (a source for the specific off-diagonal 
mean field) induces their finite value, or we choose, as an initial 
state of the self-consistent procedure, a state that requires them to be 
finite. This second case is never realized in the studies we perform in 
this manuscript. Accordingly, we will adopt only Fock terms that couple 
either the same orbitals at neighboring sites (as in the kinetic energy) 
or couple different orbitals at the same site (as in the 
orbital-polarization interaction term $\propto D$). 
These latter Fock terms will turn out to be of fundamental importance
to describe faithfully the physics of $t_{2g}$ electrons in presence
of defects. 
We remark that the HA is applied to the on-site interactions $H_{U}$ 
in a way similar to that used to implement local Coulomb interactions
within local density approximation (LDA) in the so-called LDA+$U$ 
method.\cite{Ani91}

Given the above prescriptions, the derivation of the unrestricted HF 
equations is standard and we do not present it here \textit{in 
extenso\/}; more details can be found, for instance, in Refs. 
\onlinecite{Bar00,Bar04}. Following this procedure one arrives at an 
effective one-electron HF Hamiltonian,
\begin{eqnarray}
{\cal H}_{{\rm HF}} & = & \sum_{i\alpha\sigma}\varepsilon_{i\alpha\sigma}
n_{i\alpha\sigma}+\sum_{i\alpha\mu\sigma}\beta_{i\alpha\mu\sigma}
c_{i\alpha\sigma}^{\dagger}c_{i\mu\sigma}^{},\nonumber \\
 & + & \sum_{\left\langle ij\right\rangle}
\sum_{\alpha\mu\sigma}t^{\gamma}_{i\alpha j\mu; \sigma}
c_{i\alpha\sigma}^{\dagger}c_{j\mu\sigma}^{}\,.\label{HHF}
\end{eqnarray}
This Hamiltonian can be diagonalized numerically, and the mean fields 
appearing in the parameters (see below) of the HF Hamiltonian 
(\ref{HHF}), and the HF orbitals can be determined 
self-consistently within an iterative procedure. The orbital energies 
$\varepsilon_{i\alpha\sigma}$ are defined as follows,
\begin{eqnarray}
\varepsilon_{i\alpha\sigma}\! & = & \varepsilon_{c}^{0}\delta_{c\alpha}
+U\langle n_{i\alpha\bar{\sigma}}\rangle+(U-2J_{H})\sum_{\mu\neq\alpha}
\langle n_{i\mu\bar{\sigma}}\rangle,\nonumber \\
 & + & (U-3J_{H})\sum_{\mu\neq\alpha}\langle n_{i\mu\sigma}\rangle
+\sum_{j(i)}V_{ij}\langle n_{j}\rangle
+V_{D}\sum_{\mathbf{m}\in{\cal D}}\xi_{im}\nonumber \\
 & + & \frac{1}{2}V_{ab}(\delta_{a\alpha}-\delta_{b\alpha})\sum_{j(i)\in ab}
\langle\tau_{j}^{z}\rangle\nonumber \\
 & - &  \frac{1}{2}V_{c}(\delta_{a\alpha}-\delta_{b\alpha})\sum_{j(i)\in c}
\langle\tau_{j}^{z}\rangle\,.
\label{eq:ehf}
\end{eqnarray}
We have introduced the orbital moments $\langle\tau_{j}^{z}\rangle$ for
the nearest neighbors $j$ of the considered site $i$ in the same plane 
$ab$ and along the $c$ axis, respectively, labeled $j(i)$. The parameter 
$\xi_{im}=1$ if the site $i$ belongs to a cube which surrounds a 
particular defect labeled by $m$ in Eq. (\ref{VD}), i.e., 
$i\in{\cal C}_{m}$, and $\xi_{im}=0$ otherwise. The above equations may 
be further simplified by introducing total electron densities 
$\{n_{i}\}$ and magnetizations $\{m_{i}\}$ at site $i$, and electron 
densities $\{n_{i\alpha}\}$ and magnetizations $\{m_{i\alpha}\}$ per 
orbital $\alpha$ at site $i$,
\begin{eqnarray}
n_{i} & \equiv & \sum_{\alpha\sigma}\langle n_{i\alpha\sigma}\rangle,
\label{dens}\\
m_{i} & \equiv & 
\sum_{\alpha}\langle n_{i\alpha\uparrow}-n_{i\alpha\downarrow}\rangle,
\label{mdens}\\
n_{i\alpha} & \equiv & \sum_{\sigma}\langle n_{i\alpha\sigma}\rangle,
\label{odens}\\
m_{i\alpha} & \equiv & 
\langle n_{i\alpha\uparrow}-n_{i\alpha\downarrow}\rangle.
\label{modens}
\end{eqnarray}
One finds,
\begin{eqnarray}
\varepsilon_{i\alpha\sigma} & = & \varepsilon_{c}^{0}\delta_{c\alpha}
+\left(U-\frac{5}{2}J_{H}\right)n_{i}-\frac12 (U-5J_H)n_{i\alpha}\nonumber \\
&-&\frac{1}{2}\sigma (U-J_{H})m_{i\alpha}-\frac12\sigma J_Hm_i
+  \sum_{j(i)}V_{ij}\langle n_{j}\rangle+V_{D}\xi_{i}\nonumber \\
&+&  \frac{1}{2}V_{ab}(\delta_{a\alpha}-\delta_{b\alpha})
\sum_{j(i)\in ab}\langle\tau_{j}^{z}\rangle\nonumber \\
 & - & \frac{1}{2}V_{c}(\delta_{a\alpha}-\delta_{b\alpha})
\sum_{j(i)\in c}\langle\tau_{j}^{z}\rangle.
\end{eqnarray}
Note that the on-site charge repulsive term
$U(n_{i}-\frac{1}{2}n_{i\alpha})-\frac{5}{2}J_{H}(n_{i}-n_{i\alpha})$
does not contain self-interactions. The terms that depend on magnetic
moments $\{m_{i\mu}\}$ are responsible for the magnetic order found
in the realistic regime of parameters, see Secs. V and VI. In the
weak-coupling regime, where the system is metallic, these terms give
magnetic instabilities driven by the Stoner parameter,\cite{Sto90} 
being in the present three-orbital model $I_{{\rm HF}}=U+2J_{H}$.

Near the defect at site $m$, one finds the local off-diagonal elements
of the HF Hamiltonian $\beta_{i\alpha\mu\sigma}$ at site 
$i\in{\cal C}_{m}$ given by the orbital polarization term (\ref{HD}) 
and by the Fock terms of the Coulomb interaction,
\begin{equation}
\beta_{i\alpha\mu\sigma}\!=\!\sum_{\mathbf{m}\in{\cal D}}\!\xi_{im}\!
\left\{D\lambda_{\alpha\mu}(\mathbf{r}_{i}\!-\!\mathbf{R}_{m})
-(U\!-\!3J_{H})\langle c_{i\mu\sigma}^{\dagger}
c_{i\alpha\sigma}^{}\rangle\right\},
\label{beta}
\end{equation}
where the parameters $\{\xi_{im}\}$ are defined as in Eq. (\ref{eq:ehf}).
As we show in Secs. III and IV, the second term in Eq. (\ref{beta}) 
is crucial for $D\neq0$
as it renormalizes the orbital mixing term $\propto D$ and makes it 
possible to find the orbitals that optimize the energy of the system.
The hopping parameters $t_{i\alpha j\mu\sigma}$ are renormalized
by the Fock term that stem from the intersite Coulomb interaction
(\ref{Vee}),
\begin{equation}
t_{i\alpha j\mu\sigma}=t^{\gamma}_{\alpha\mu}
-V_1\langle c_{j\mu\sigma}^{\dagger}c_{i\alpha\sigma}^{}\rangle,
\label{teff}
\end{equation}
and this renormalization is, for instance, responsible for the different
bandwidths of the minority and the majority bands in transition metals.
\cite{Bar04} According to the nearest-neighbor nature of the ``source'' 
term (the kinetic energy), also the related Fock terms stem from 
the nearest-neighbor Coulomb interaction matrix element $V_1$. 
As a matter of fact, as discussed above,
all other Fock terms will simply vanish.

\subsection{Undoped Mott-Hubbard insulator}
\label{sec:MIT}

We begin with discussing the reference electronic structure of the
two phases with broken symmetry in spin and orbital space relevant
to YVO$_{3}$: the $CG$ phase and the $GC$ phase. The spin and orbital
order in these phases is shown schematically in Figs. \ref{fig:1}(b)
and \ref{fig:1}(c), respectively. Unrestricted HF calculations
have been performed for an $8\times8\times8$ supercell with PBC; larger 
systems have also been considered. In the case of an undoped system, a 
smaller supercell like $6\times6\times6$ or even $4\times4\times4$ would 
suffice as the HF convergence is indeed very fast and the final results 
for charge, spin and orbital density distributions as well as for the 
total and orbital resolved density of states (DOS) do not depend on 
system size, in both symmetry-broken
phases, for sufficiently large $N$ ($N\geq4$). Within our numerical
studies, we have used several parameters which are considered realistic 
for YVO$_{3}$, but here we present only the representative results 
obtained for two parameter sets given in Table \ref{tab:para}.

\begin{table}[b!]
\caption{Standard parameter sets used in the numerical calculations;
all parameters are given in eV. Defect potential $V_{\rm D}$ (\ref{VD})
contributes only in doped systems.} 
\vskip .2cm
\begin{ruledtabular}
\begin{tabular}{cccccccc}
set & $t$ & $\varepsilon_c^0$ & $U$ & $J_H$ & $V_{ab}$ & $V_c$ & $V_{\rm D}$\cr 
\colrule
A   & 0.0 &   0.1             & 4.0 &  0.6  &   0.03   & 0.05  & 1.0 \cr
B   & 0.2 &   0.1             & 4.0 &  0.6  &   0.03   & 0.05  & 1.0 \cr
\end{tabular}
\end{ruledtabular}
\label{tab:para}
\end{table}

In the relevant regime of parameters for the $R$VO$_{3}$ perovskites,
the electrons in undoped YVO$_{3}$ are localized as in a Mott insulator.
We consider here the same set of parameters as used previously in
Refs. \onlinecite{Ole07,Hor11}: $t=0.2$ eV, $\varepsilon_{c}^{0}=0.1$
eV, $U=4$ eV and $J_{H}=0.6$ eV. The value of $t=0.2$ eV has been
estimated using the charge transfer model. For Hund's exchange $J_{H}$,
we have adopted a value that is somewhat screened with respect to
its atomic value $J_{H}^{{\rm atom}}=0.64$ eV.\cite{Ole07} Finally,
the value of $U$ have been selected in order to obtain a value for
the superexchange parameter $J=4t^{2}/U=40$ meV, which is consistent
with the results obtained in neutron scattering experiments for spin
excitations.\cite{Ulr03} As regards the JT parameters $V_{ab}$
and $V_{c}$, we have chosen values that are in the expected range for 
LaVO$_{3}$. 

The value of the CF potential
$\varepsilon_{c}^{0}=0.1$ eV guarantees that, in absence of the 
orbital-polarization interaction $D$, the $c$ orbital is occupied at 
each V$^{3+}$ ion. In this parameter range, and as long as any defect is 
absent, one may simplify the calculation of the electronic structure in 
the HF approximation as the spin of the $c$ electron determines the 
magnetic moment at each vanadium ion, which is in the high-spin (HS) 
state ($S=1$) due to finite Hund's exchange. In this respect, the HF 
calculations reduce to those performed within the simplified two-flavor 
model.\cite{Hor11} Depending on the starting initial conditions, 
one finds two locally stable configurations, the $CG$ and the $GC$ 
phases shown schematically in Figs. \ref{fig:1}(a) and \ref{fig:1}(b), 
with a uniformly distributed charge of two electrons at each vanadium ion.

\begin{figure}
\includegraphics[width=8cm]{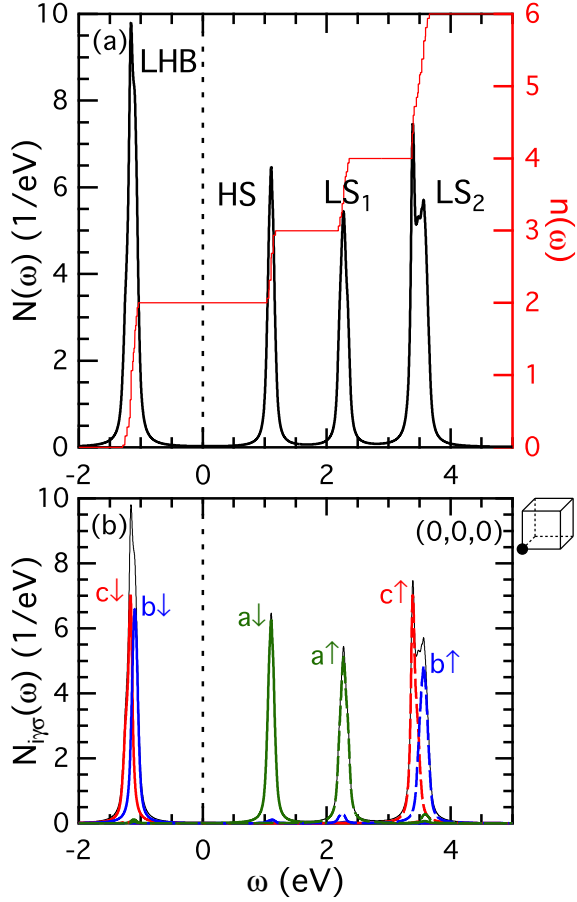} \caption{(Color online) Densities
of states for the Mott insulator in the $CG$ phase calculated for an 
$8\times8\times8$ periodic cluster:
(a) total DOS $N(\omega)$ (left scale) and the total electron filling
$n(\omega)$ (right scale); 
(b) local spin-orbital DOSs $N_{i\alpha\sigma}(\omega)$ at the atom at 
site $i=(0,0,0)$, orbital $\alpha$ and spin $\sigma$. 
The spectra reflect the spin-orbital order at (0,0,0) and
$\downarrow$-spin ($\uparrow$-spin) states are drawn by solid (dashed)
lines, respectively. Vertical dashed line indicates the 
Fermi energy at $\omega=0$; parameters as in set B in 
Table \ref{tab:para}.}
\label{fig:2}
\end{figure}

\begin{figure}[t!]
\includegraphics[width=8cm]{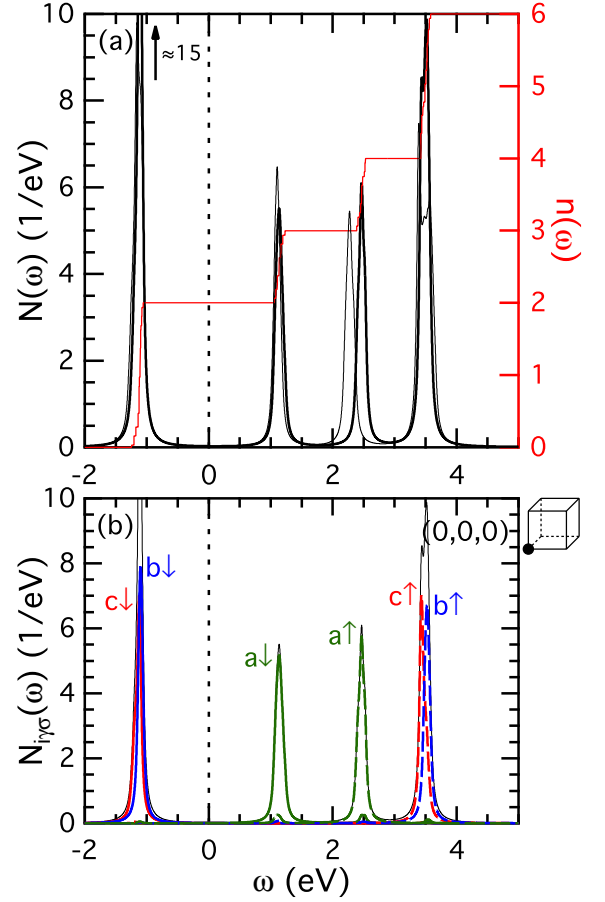}
\caption{(Color online) Densities of states as obtained for the Mott 
insulator in the $GC$ phase for an $8\times8\times8$ system: 
(a) total DOS $N(\omega)$ (left scale, heavy line), the same for the
$CG$ phase (thin line) shown for comparison, and the total electron 
filling $n(\omega)$ (right scale); 
(b) local spin-orbital DOSs $N_{i\alpha\sigma}(\omega)$ at site (0,0,0), 
orbital $\alpha$ and spin $\sigma$. Here  $\downarrow$-spin 
($\uparrow$-spin) states are shown by solid (dashed) lines, 
respectively. Fermi energy and parameters as in Fig. 2.}
\label{fig:3}
\end{figure}

The current self-consistent unrestricted HF calculations, performed in
the undoped regime, give the electronic structures of the $CG$ and the
$GC$ phase,\cite{notew} shown in Figs. \ref{fig:2} and \ref{fig:3}.
For the present parameters $t_{2g}$ electrons are fairly 
well localized (as $t\ll U$) and 
a broken-symmetry ground state is found. In such a case, the HF approach 
is rather successful and provides not only a good description of the 
occupied states, but also a quite satisfatory description of the excited 
ones. In fact, in the present case also the structure of the upper
Hubbard band (UHB) is quite well reproduced. To demonstrate this, let us 
consider first the $CG$ phase. At a representative atom in position 
(0,0,0), see Fig. \ref{fig:1}(b), one finds the DOS with well separated 
occupied states in the LHB and with empty states in the UHB, which 
consists of three peaks.
The almost-classical ground state corresponds to an occupied wave
function $|c\downarrow b\downarrow\rangle$. While the $c$ electron
is perfectly localized ($n_{c}\simeq0.993$), the occupancy of the $b$ 
orbital is somewhat lower than one, $n_{b}\simeq0.977$. This is due 
to weak fluctuations of the orbital flavor between nearest-neighbor
sites along the $c$ axis, where all spins are aligned ($C$-AF) and
orbitals alternate ($G$-AO). Consequently, the $|a\downarrow\rangle$
state has finite low electron density ($n_{a}\simeq0.016$).

The unoccupied part of electronic structure, the UHB, consists of three 
distinct maxima, see Fig. \ref{fig:2}(a). The lowest one corresponds to 
the HS excitation and the system is a Mott insulator with a large MH 
gap: $\Delta_{{\rm MH}}\simeq2.20$ eV. This value is obtained as the 
difference between the energies measured at the center of gravity (the 
first moment) of the peaks corresponding to the highest occupied and the 
lowest unoccupied HF orbitals: $|b\downarrow\rangle$ and 
$|a\downarrow\rangle$, respectively, see Fig. \ref{fig:2}(b). Note that 
the value of the MH gap is given here by the energy of the HS charge 
excitation $d_{i}^{2}d_{j}^{2}\rightleftharpoons d_{i}^{3}d_{j}^{1}$
to the $S=3/2$ spin state,
\begin{equation}
E_{{\rm HS}}=U-3J_{H}\,,\label{hs}
\end{equation}
with a $a^{1}b^{1}c^{1}$ configuration of $\downarrow$-spin electrons
in the fully localized (atomic) limit.\cite{Ole05} 

The UHB contains also low-spin (LS) excited states. As pointed out 
before,\cite{Hor11}, their excitation energies obtained in the HF 
approximation (at $\varepsilon_{c}=0$),
\begin{eqnarray}
E_{{\rm LS1}} & \simeq& U-J_{H}\,,\label{ls1}\\
E_{{\rm LS2}} & \simeq& U+J_{H}\,.\label{ls2}
\end{eqnarray}
are systematically lower by $J_{H}$ than the exact ones and the upper
state is doubly degenerate instead of the lower one, as the 
quantum-fluctuation driven processes (spin-flip and ``pair-hopping'' 
terms) are neglected.\cite{notex} For the considered ground state 
$|c\downarrow b\downarrow\rangle$, one finds first a LS ($S=\frac12$) 
state with the $|a\uparrow\rangle$ state occupied, see Fig. 
\ref{fig:2}(b). The energy of this excitation, measured from the energy 
of the highest occupied $|b\downarrow\rangle$ state, is 
$E_{{\rm LS1}}=3.37$ eV, which is indeed very close to the value 
obtained from Eq. (\ref{ls1}) in the atomic limit, 
$\left(U-J_{H}\right)=3.4$ eV. The two remaining LS states have energies 
$E_{{\rm LS2}}^{(c)}=4.47$ eV and $E_{{\rm LS2}}^{(b)}=4.64$ eV, 
corresponding to double occupancies in the $c$ and $b$ orbitals, 
respectively. These states have been labeled with the added electron to 
the ground state, $|c\uparrow\rangle$ and $|b\uparrow\rangle$ 
in Fig. \ref{fig:2}(b). The above energies are again
very close to the atomic limit values, $\varepsilon_{c}+(U+J_{H})=4.5$
eV and $(U+J_{H})=4.6$ eV, respectively.

The maxima in the total DOS $N(\omega)$ are well separated from
each other. This can be clearly seen in the total electron filling
up to energy $\omega$,
\begin{equation}
n(\omega)=\int_{-\infty}^{\omega}d\omega' N(\omega')\,,
\label{nw}
\end{equation}
which is almost constant in between the maxima corresponding to 
different states, see Fig. \ref{fig:2}(a). In particular, no states can 
be  found within the MH gap.

The overall picture obtained for the $GC$ phase is very similar, again
with the LHB accompanied by three peaks in the UHB (see Fig. 
\ref{fig:3}). In this case, the $G$-AF ground state makes the electrons 
even more localized and one finds (down-spin) electron densities 
$n_{c}\simeq0.993$, $n_{b}\simeq0.989$ and $n_{a}\simeq0$ at the 
representative (0,0,0) atom, see Fig. \ref{fig:1}(c). The system is a 
good MH insulator with total electron filling $n(0)=2$. A better 
electron localization than in the $CG$ phase may also be concluded from 
a somewhat increased value of the MH gap, $\Delta_{{\rm MH}}\simeq2.24$ 
eV. The low-spin excitations are found now at energies: 
$E_{{\rm LS1}}=3.57$ eV, 
$E_{{\rm LS2}}^{(c)}=4.53$ eV and $E_{{\rm LS2}}^{(b)}=4.62$ eV. The 
largest difference with respect to the $CG$ phase of  $0.2$ eV is found 
for the LS1 excitation energy $E_{{\rm LS1}}$, see Fig. \ref{fig:3}(b). 
We suggest that this increased excitation energy originates from a  
highly localized $(a\uparrow)$ component within the $ab$ plane 
(in the $GC$ phase) as the very same orbital is stronger localized here 
at the neighboring sites along $a$ and $b$ directions than in the $CG$ 
phase. We also note the increased height of the peaks which indicates 
again stronger localization.

In summary, we have shown by the HF analysis of both $CG$ and $GC$
phases that the DOS consists of the LHB and the UHB, and that the latter 
has a well defined internal structure of excited states. The MH gaps are 
very similar in both magnetic structures for the chosen set of 
parameters. In the next sections, we analyze the change of local 
electronic state near charged defects introduced by doping and show how 
such defects change the above idealized electronic structure.

\section{Optimized atomic orbitals}
\label{sec:atom}

\begin{figure}[b!]
\includegraphics[width=8.2cm]{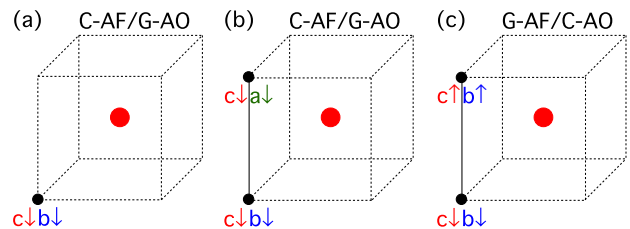} 
\caption{(Color online) 
Reference electronic configurations considered in the reference ground 
state at $D=0$ in Secs. \ref{sec:atom} and \ref{sec:bond} for: 
(a) a single atom (dot) in the $CG$ phase; 
(b) a bond (solid line connecting two dots) along the $c$ axis in the 
$CG$ phase; 
(c) a bond (solid line connecting two dots) along the $c$ axis in the 
$GC$ phase. 
In each case, the state $|c\downarrow b\downarrow\rangle$ is selected 
as the initial ground state at the (0,0,0) atom. The relative position 
of the defect state in the center of each cube is marked by a filled 
(red) sphere.}
\label{fig:4}
\end{figure}

To develop systematic understanding of the evolution of the electronic
structure under increasing doping, it is helpful to analyze first the
atomic problem for a representative atom near a Ca defect embedded in 
the $CG$ phase of YVO$_{3}$. Thus, we consider the atomic limit, i.e., 
no hopping ($t=0$), as in the parameter set A of Table I. The HF mean 
field terms act on the electronic states of such an embedded atom with 
the ground state $|c\downarrow b\downarrow\rangle$, see Fig. 
\ref{fig:4}(a).
This problem is solved self-consistently in the HF approximation
described in the previous section and the solution is systematically
compared to the ED results obtained for the very same external mean 
fields.

Although the electronic configuration $d^{2}$ at an undoped V$^{3+}$ ion 
in the vicinity of the defect does not change with respect to the Mott
insulator described in Sec. \ref{sec:MIT}, the occupied orbitals are now 
modified due to the electron polarization term Eq. (\ref{HD}).
Here we focus on the description of a single atom and compare the HF
results with those of the ED with the aim at validating the scheme that 
will be used in Secs. V and VI for the bulk. The results are mainly 
presented for varying polarization parameter $D$, responsible for the 
adjustment of the orbitals to the field which acts on them.

\begin{figure}[t!]
\includegraphics[width=7.8cm]{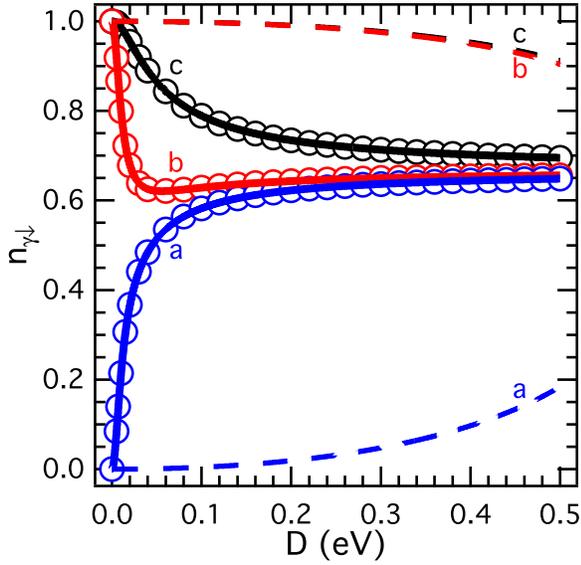} 
\caption{(Color online) 
Electron densities $n_{\gamma\downarrow}$ in the pristine orbitals 
$\gamma\in\{a,b,c\}$ for an atom embedded in the $CG$ phase, see Fig. 
\ref{fig:4}(a), as functions of increasing orbital polarization $D$. 
Solid lines from HF, dashed lines from HA and hollow circles from ED. 
Parameters as in set A of Table \ref{tab:para}.}
\label{fig:5}
\end{figure}

\begin{figure}[t!]
\includegraphics[width=7.8cm]{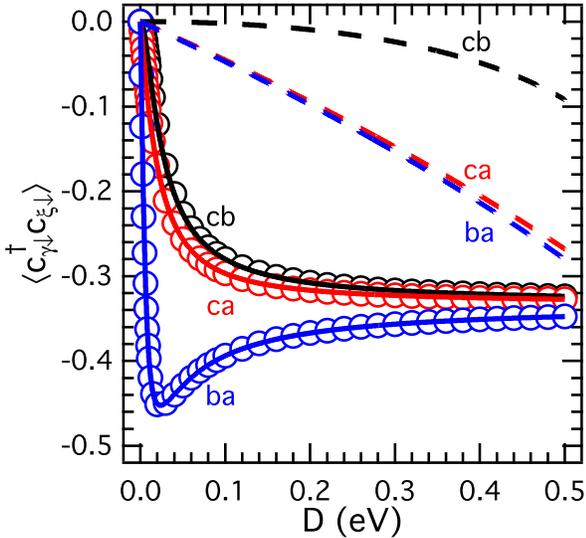} 
\caption{(Color online) Off-diagonal elements 
$\left\langle c_{\gamma\downarrow}^{\dagger}c_{\xi\downarrow}\right\rangle$
($\gamma\neq\xi$) of the density matrix for spin-down pristine orbitals
for an atom embedded in the $CG$ phase, see Fig. \ref{fig:4}(a),
as functions of increasing orbital polarization $D$. Solid lines
from HF, dashed lines from HA and hollow circles from ED. 
Parameters as in set A of Table \ref{tab:para}.}
\label{fig:6}
\end{figure}

The set of $t_{2g}$ orbitals $\{a,b,c\}$ Eq. (\ref{frozen}), also called
\textit{pristine\/} orbitals hereafter, is selected as orthogonal
basis in the undoped YVO$_{3}$ by lattice distortions, which lower
the $c$ orbitals by a finite CF energy $\varepsilon_{c}^{0}$.
\textit{A priori\/}, $t_{2g}$ orbitals are expected to adjust to the 
orbital polarization field $D$ in the vicinity of a defect and a
new set of orthogonal orbitals will become the optimal basis for the 
embedded atom. We have determined this optimal basis using the 
self-consistent HF calculations described in the previous section and 
found that all three pristine $t_{2g}$ orbitals $\{a,b,c\}$ are strongly 
modified; already for $D=0.1$ eV, the 
occupied orbitals look quite different from $\{a,b,c\}$ ones, see below. 
This gradual orbital rotation can be easily recognized by analyzing the 
electron densities, 
$n_{\gamma\downarrow}\equiv\langle n_{\gamma\downarrow}\rangle$,
and the off-diagonal elements of the density matrix, 
$\langle c_{\gamma\downarrow}^{\dagger}c_{\xi\downarrow}^{}\rangle$,
shown in Figs. \ref{fig:5} and \ref{fig:6}, respectively, for the
$CG$ phase. 

The electron densities are $n_{c\downarrow}=n_{b\downarrow}=1$ and 
$n_{a\downarrow}=0$ for the reference state at $D=0$. One finds that 
they change fast with increasing $D$ for the orbitals $a$ and $b$ (Fig. 
\ref{fig:5}), while the decrease in the electron density within the $c$ 
orbital is slower. Therefore, in the weak polarization regime of 
$D<0.02$ eV, the two-flavor model\cite{Hor11} is sufficient. On the 
contrary, for $D\simeq 0.1$ eV, all three orbitals contribute to the
ground state with similar electron densities --- the electron densities 
in the $a$ and $b$ orbitals are almost equal and $n_{c\downarrow}<0.8$. 
Certainly, the two orbitals occupied in this case are quite different 
from the pristine $c$ orbital and one of the doublet $\{a,b\}$ orbitals 
and the full orbital space has to be considered. This observation is 
further supported by the values of the off-diagonal matrix elements 
(Fig. \ref{fig:6}), which become similar to each other for $D\simeq0.1$ 
eV, while for $D<0.02$ eV the orbital mixing is large only between the 
$a$ and $b$ orbitals. For the largest value of $D=0.5$ eV studied here, 
all three $t_{2g}$ orbitals are almost equally occupied (Fig. 
\ref{fig:5}) and the off-diagonal orbital elements are almost the same 
for each pair (Fig. \ref{fig:6}).

It is very remarkable that the HF results presented in Figs. \ref{fig:5}
and \ref{fig:6} agree exactly with the ED results. This shows that 
the quantum fluctuations, going beyond HF approximation, are irrelevant 
for the ground state in the present regime of parameters. In fact, the 
electron localization and the symmetry breaking in the spin and orbital 
space, which follow from the large value of $U$, suppress quantum 
fluctuations around the exact state with two occupied optimal orbitals. 
In contrast, the results obtained within the simpler HA (without Fock
orbital terms), turn out to be completely unsatisfactory and this  
approximation gives an unrealistic description of the orbital states, 
see Figs. \ref{fig:5} and \ref{fig:6}. Within HA, the orbitals $b$ and 
$c$ are almost unchanged and remain still the occupied orbitals even for 
relatively large values of $D>0.1$ eV, where also their rotation 
(mixing) hardly occurs. This demonstrates that the Fock terms are 
essential in the present problem, as these off-diagonal terms are 
responsible for a gradual adjustment of the orbital subsystem to the 
orbital polarization term Eq. (\ref{HD}) which drives the orbital mixing.

\begin{figure}[t!]
\includegraphics[width=7.8cm]{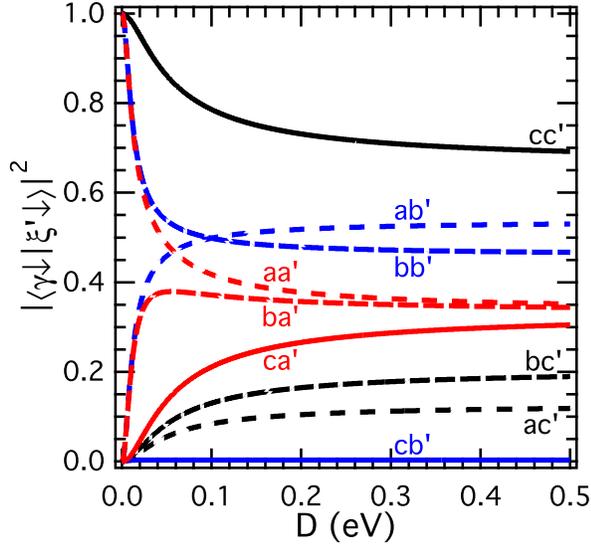} \caption{(Color online) 
Fractions of pristine orbitals $\{|\gamma\rangle\}$ in the rotated 
orbitals $\{|\xi'\rangle\}$ for an atom embedded in the $CG$ phase, 
see Fig. \ref{fig:4}(a), as functions of increasing orbital 
polarization $D$. Parameters as in set A of Table \ref{tab:para}.}
\label{fig:7}
\end{figure}

The gradual evolution of the two occupied rotated orbitals 
$\{|b'\rangle,|c'\rangle\}$ (and the unoccupied orbital 
$\{|a'\rangle\}$) relative to the original $t_{2g}$ basis as function of 
the orbital polarization field $D$ is illustrated by the overlap
functions $\langle\gamma\downarrow|\xi'\downarrow\rangle$ in Fig. 
\ref{fig:7}. The initial state with occupied orbitals 
$\{|b\rangle,|c\rangle\}$ changes gradually with increasing $D$ and the 
occupied orbitals $\{|b'\rangle,|c'\rangle\}$ are linear combinations 
of the pristine ones. One finds once more that, except for the region 
of rather small $D<0.02$ eV, all three pristine orbitals contribute and
the orbitals are strongly modified. The new states $|c'\rangle$ and 
$|a'\rangle$ arise as linear combinations of all three $t_{2g}$ orbitals 
$\{a,b,c\}$. In contrast, the occupied orbital state $|b'\rangle$
has no component with $|c\rangle$ orbital character and is just a
linear combination of $\{a,b\}$ orbitals. By considering solely this
orbital state, one could therefore justify \textit{a posteriori\/}
the two-orbital model,\cite{Hor11} but we note that: 
(i) the actual ground state involves the occupied $|c'\rangle$ orbital 
too, which has a significantly modified shape with respect to the 
original $|c\rangle$ state, and 
(ii) the presence of finite kinetic energy between such orbital states
will induce a further redistribution of the pristine orbital character 
over the two occupied orbitals $\{|b'\rangle,|c'\rangle\}$, and will 
also modify the $|a'\rangle$ orbital contributing to excited states.
The difference found here between the $a$ and $b$ orbital character
in the occupied $|c'\rangle$ state follows from the mean fields that
arise due to the JT interactions.

\begin{figure}[t!]
\includegraphics[width=8.2cm]{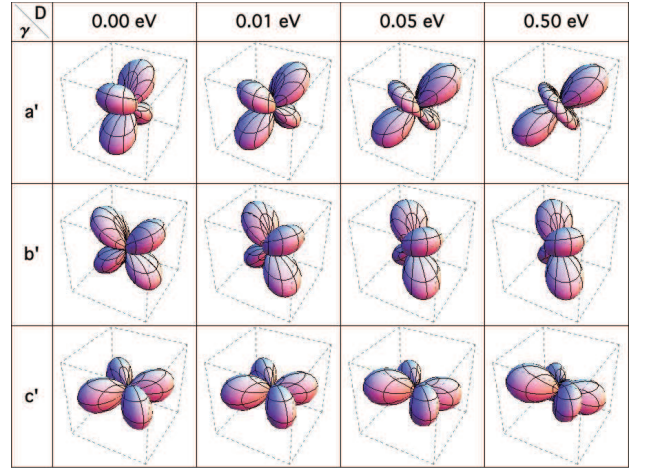} 
\caption{(Color online) 
Optimal orbital basis for an embedded atom in the $CG$ phase at 
position (0,0,0), see Fig. \ref{fig:4}(a), for increasing orbital 
polarization field $D$. The defect which is the source of the orbital 
polarization term is located at position $(\frac12,\frac12,\frac12)$. 
The three rotated orbitals are labeled by $\gamma=a',b',c'$, with the 
orbitals $b'$ and $c'$ occupied in the ground state. At $D=0$ eV 
(first column) $\{a',b',c'\}$ orbital coincide with $\{a,b,c\}$.
Parameters as in set A of Table \ref{tab:para}.}
\label{fig:8}
\end{figure}

The local orbital basis gradually rotates and adjusts itself to the
orbital polarization when $D$ increases, as shown in Fig. \ref{fig:8}.
The Coulomb repulsion between the exceeding electron at the Ca defect in 
the center of the V$_8$ cube, see Fig. \ref{fig:4}(a), and the V$^{3+}$ 
ion adjusts the shape of the occupied orbitals, 
$\{|b'\rangle,|c'\rangle\}$, in such a way that the electrons involved 
increase their distance from each other, and not too much of the JT and
CF energy is lost. Therefore, the orbital shapes become all nonequivalent
for $D>0$ and evolve gradually into two geometrically similar occupied
orbitals $\{b',c'\}$, with $c'$ practically lying in the plane 
perpendicular to the direction along which the defect resides, and the 
empty cigar-like $a'$ orbital, directed towards the defect site. This 
final configuration can be easily recognized at the largest studied 
value $D=0.5$ eV. In the large D regime, the rotated wave functions are:
\begin{eqnarray}
|a'\rangle &=& \frac{1}{\sqrt{3}}( |a\rangle +|b\rangle +|c\rangle ), \\
|b'\rangle &=& \frac{1}{\sqrt{2}}(-|a\rangle +|b\rangle), \\
|c'\rangle &=& \frac{1}{\sqrt{6}}(-|a\rangle -|b\rangle +2|c\rangle).  
\end{eqnarray}
In a doped system a hole will go into the topmost occupied orbital 
$|b'\rangle$, which is an odd linear combination of $|a\rangle$ and  
$|b\rangle$, and does not contain the third $c$ flavor.

When $D$ is finite and the CF term as well as and the mean field terms 
acting on the atom are neglected, the local problem given by the 
orbital polarization maps onto the hole hopping problem on a triangle, 
with the energy spectrum: $\{-D,-D,+2D\}$. Having two electrons at 
the V$^{3+}$ ion, the two degenerate states with energy $-D$ are then
occupied. The increasing similarity of the occupied states anticipated
from this result is indeed observed when $D$ increases in presence of 
other terms. In between $D=0$ and the largest value $D=0.5$ eV studied 
here, one can distinguish two qualitatively different regimes: 
(i) for small $D\lesssim0.02$ eV only the orbitals $\{a,b\}$ mix, while 
the $c$ orbital still does not change as it is stabilized by the 
CF energy $\varepsilon_{c}^{0}=0.1$ eV; 
(ii) when $D$ becomes of the order of $\varepsilon_{c}^{0}$,
the $c$ orbital is destabilized and all three orbitals rotate further
towards their final shapes. The case of $D\sim0.05$ eV shown in Fig.
\ref{fig:8} gives orbitals that are already close to those found
at the largest studied value $D=0.5$ eV. Note also that weak JT 
interactions counteract the orbital rotation, but the orbital 
polarization is the dominating term for the present set of parameters, 
and the basis rotation is practically completed already for $D\simeq0.1$ 
eV (see Fig. \ref{fig:7}).

\begin{figure}[t!]
\includegraphics[width=7.8cm]{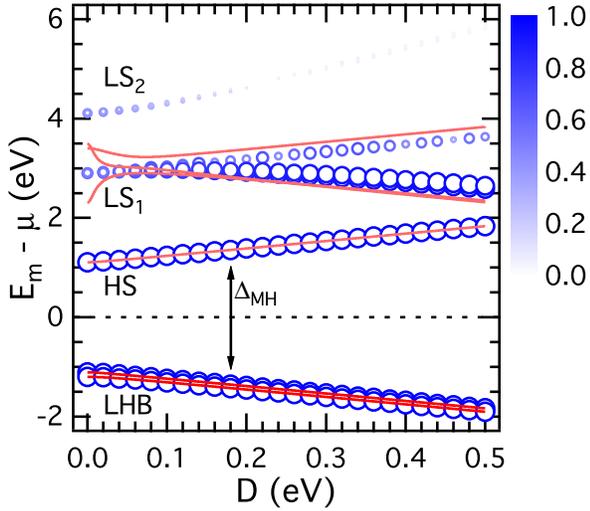} 
\caption{(Color online) 
Excitation energies $E_{m}$ for an atom embedded in the $CG$ phase, 
see Fig. \ref{fig:4}(a), measured with respect to the chemical 
potential, as functions of increasing orbital polarization $D$. 
Solid lines from HF (heavy/light lines for occupied/unoccupied
states); and hollow circles from ED (the color scale on the right
indicates the spectral weight of each level, as well as the size of
the symbols). The MH gap is found between the LHB and the HS state
of the UHB; the labels of the LS excited states refer only to the 
regime $D\simeq0$. The chemical potential $\mu$ is fixed at the 
center of the MH gap $\Delta_{{\rm MH}}$. 
Parameters as in set A of Table \ref{tab:para}.}
\label{fig:9}
\end{figure}

Orbital polarization influences also the one-particle excitations of an 
atom close to a defect. Given the ground state energy $E_{0}(d^{n})$ of 
a Mott insulator in the $d^{n}$ electronic configuration, these
excitations in correlated insulators are calculated as follows: \\
 --- for exciting a hole, as in photoemission (PES) spectra,
\begin{equation}
E_{m}=E_{m}(d^{n-1})-E_{0}(d^{n}),
\label{exhole}
\end{equation}
--- and similarly for adding an electron, corresponding to inverse
photoemission (IPES) spectra,\cite{Mei93}
\begin{equation}
E_{m}=E_{m}(d^{n+1})-E_{0}(d^{n}).
\label{exelec}
\end{equation}
We determined the excitation energies for an atom embedded in the $CG$ 
phase in proximity of a defect using the unrestricted HF approximation 
and compared them with the corresponding ED results, as shown in Fig. 
\ref{fig:9}. For the sake of convenience, the chemical potential $\mu$, 
fixed as usual in the center of the MH gap, is subtracted in each case.

\begin{figure}[t!]
\includegraphics[width=7.8cm]{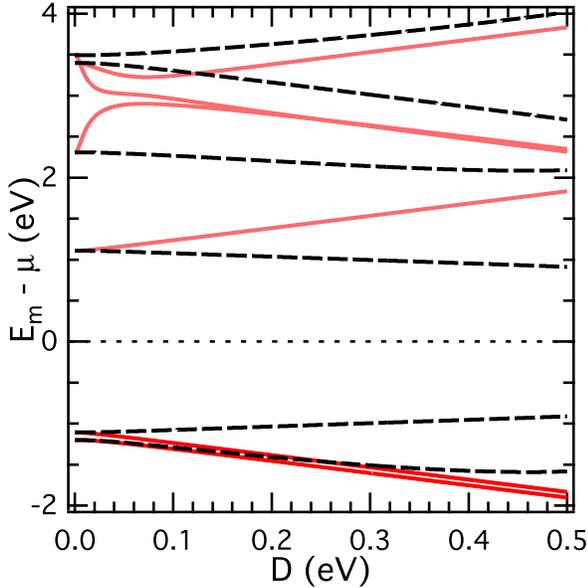} \caption{(Color online) 
Excitation energies $E_{m}$ for an atom embedded in the $CG$ phase, 
see Fig. \ref{fig:4}(a), obtained in the HF (solid lines) and in the HA 
(dashed lines) for increasing orbital polarization $D$. The energies 
$E_{m}$ are measured with respect to the chemical potential $\mu$ 
(dotted line). Parameters as in set A in Table \ref{tab:para}.}
\label{fig:10}
\end{figure}

In the HF spectra one finds two PES excitations that correspond
to removing an electron either from $b'$ or $c'$ orbitals; the energy 
difference between them is small and comes from the CF splitting 
$\varepsilon_{c}^{0}$. These states form the occupied LHB. The apparent 
decrease in energy present in Fig. \ref{fig:9}, 
$({\partial E_{m}}/{\partial D})\cong-1.5$ instead of the expected 
$({\partial E_{m}}/{\partial D})\cong-1$, follows 
from the subtraction of the chemical potential. The four excitations at 
positive energies correspond to the four possible $d^{3}$ excited states
analyzed for the undoped case in Sec. \ref{sec:MIT}: 
(i) the HS state has the lowest energy and the $|a\downarrow\rangle$ 
state is occupied, 
(ii-iv) in the three LS states with higher energies a $\uparrow$-spin 
state is occupied in one of the three orbitals. 
Notably, these excitation energies depend rather strongly on the orbital 
polarization $D$. The energy of the HS state ($S^{z}=-\frac{3}{2}$) 
increases with increasing $D$ when all three orbital flavors are 
occupied with the same spin, and 
$({\partial E_{m}}/{\partial D})\cong2$ before chemical potential
subtraction. As a matter of fact, such increase concerns also the 
LS2 excited state with $S^{z}=-\frac{1}{2}$ and 
$|c'\uparrow\rangle$ state occupied. On the contrary, the remaining two 
LS states of the UHB have either $(a'\uparrow)$ or 
$(b'\uparrow)$ occupations by the $\uparrow$-spin and their 
energies decrease. Note that the energies of the HS and of these latter 
LS excited states cross for a value of $D>0.5$ eV, but we have estimated
that so large values of $D$ are unrealistic.

A comparison of the HF results with the exact results found using ED 
for an embedded atom close to a defect gives an excellent agreement
for the LHB and even for the lowest HS excited state of the UHB. This
may be expected as these three states have no quantum corrections.
On the contrary, as already pointed out in Sec. \ref{sec:HF}, in the LS 
sector one finds that the excitation energies for $D=0$ are lower by 
$\sim J_{H}$ than the exact values.\cite{Hor11} When $D$ increases, the
spectral weight moves from the high-energy LS2 state with 
$E_{m}-\mu>4$ eV to the lower energy excitations with $E_{m}-\mu\sim3$ 
eV. This is in contrast to the HF results, where the one-electron wave 
functions do not allow for any spectral weight transfer, and instead 
the excitation energies change and interpolate between the relevant 
energies in a continuous way. Already for $D\simeq0.15$ eV, one finds 
that the spectral weight found in the ED has moved completely to the 
lower energy regime characteristic of LS1 states, and when $D$ increases 
further there is one excitation with large spectral weight that follows 
roughly the two lower HF excitations of LS character, and the third one 
with increasing energy, similar to the one found in the HF, see above. 
Rather small energy difference between the ED energies and the HF ones 
found here demonstrates that the quantum effects are quenched in the 
orbital system when the orbital polarization $D$ is sufficiently large. 
Even more importantly, a satisfactory agreement
between the energies found in the HF approximation and the exact values
in the LS sector shows that the HF states correctly adjust themselves
to the underlying interactions and simulate the actual multi-electron
states in a very realistic way.

One may wonder again whether the HA would not give at least a 
satisfactory description of the excitations being close to the MH gap, 
as we have shown above that for all of them the quantum corrections to
the HF energies are negligible. Instead, the HA fails also here in a 
rather spectacular way --- one finds that the MH gap exhibits a 
qualitatively incorrect behavior and decreases with increasing orbital
polarization $D$, and even the structure of the LHB, with two PES 
excitations of almost equal energies, is not correctly reproduced, 
see Fig. \ref{fig:10}.
Large systematic deviations are also found for the higher LS excited
states. All these misleading features follow from the incorrect local
basis rotation with increasing orbital polarization $D$, and thus
from the incorrect evolution of the orbital electron densities shown
in Fig. \ref{fig:5}. It can be concluded that, at least in case of
conflicting external fields (e.g. the CF, the JT interaction and the 
orbital polarization), the HA is totally unsatosfactory.

\section{A bond near the defect state}
\label{sec:bond}

The electronic structure in the vicinity of defects depends also on the 
kinetic energy (\ref{ht}). While in the undoped bulk (Sec. 
\ref{sec:MIT}), the kinetic energy is diagonal and couples two atomic 
$t_{2g}$ states with the same orbital flavor at neighboring sites, this 
is not longer the case close to defects where the local orbital basis 
has to adjust to the actual fields acting on each atom, see Sec.
\ref{sec:atom}. Therefore, one expects that qualitative changes may 
arise with respect to the case of an embedded atom when electron hopping
contributes. The analysis of an embedded bond presented in this section
addresses this issue and serves to deepen and improve the understanding 
of the HF results presented for the bulk in the low doping regime, 
analyzed in Secs. V and VI.

\subsection{Electron densities induced by orbital polarization}
\label{sec:bondd}

\begin{figure}[t!]
\includegraphics[width=7.8cm]{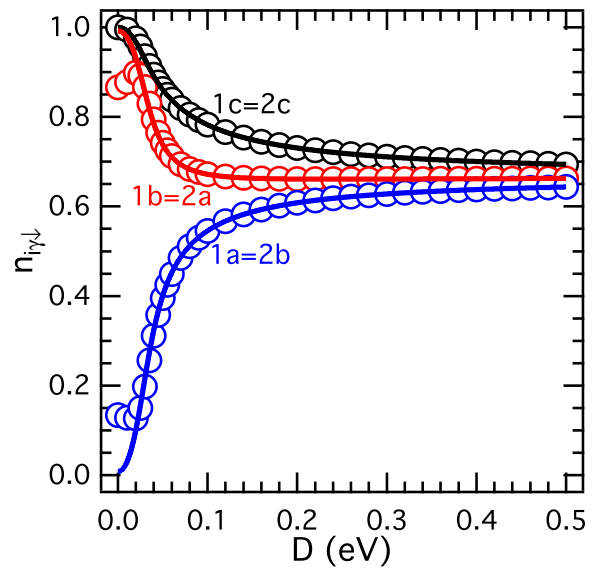}
\caption{(Color online) 
Occupations of pristine orbitals $n_{i\gamma\downarrow}$ at the 
embedded bond $\langle ij\rangle$ along the $c$ axis in the $CG$ 
phase, see Fig. \ref{fig:4}(b), as functions of increasing orbital 
polarization $D$. Solid lines from HF and hollow circles from ED. 
Parameters as in set B in Table \ref{tab:para}.}
\label{fig:11}
\end{figure}

We consider first a representative bond $\langle ij\rangle$ in the $CG$ 
phase, shown in Fig. \ref{fig:4}(b). The $c$ orbitals are occupied at 
both sites of the bond (in the absence of defects, i.e., at $D=0$) and 
the JT interactions (\ref{HJT}) favor the $b$ ($a$) orbital occupation 
at the lower (upper) site. This case is the most interesting one as the 
spin order is FM and the two electrons in the $\{a,b\}$ orbital doublet 
can delocalize along the bond. The kinetic energy is strongly suppressed 
by large local Coulomb interactions, but still one finds, within the ED, 
that orbital fluctuations along the bond contribute and modify 
the electron densities $n_{i\gamma\downarrow}$ for $\gamma=a,b$ at 
$D=0$, see Fig. \ref{fig:11}. Such fluctuations are very important 
and support the FM coupling along the vertical bonds in the undoped
material.\cite{Kha01} As we show below, the HF approximation may be still
used although it is unable to treat orbitally entangled states,
\cite{Ole06,Ole12} as these fluctuations are quenched by external fields, 
including orbital polarization $D>0$. In 
contrast, no fluctuations occur for the localized $c$ electrons and 
their density is perfectly reproduced by the HF approximation.

As in the case of a single atom analyzed in Sec. \ref{sec:atom}, an 
increasing orbital polarization $D$ modifies the occupied orbitals and 
thus the electron densities, see Fig. \ref{fig:11}. Full symmetry in 
the orbital occupancies is preserved, provided one interchanges the 
orbital flavors $\{a,b\}\leftrightarrow\{b,a\}$ between the sites $i=1$
and $j=2$. Similarly as in the embedded-atom case, the orbitals of the 
doublet active along the $c$ axis mix easier and this mixing competes 
with the JT effective interactions, which act on them at each site. The 
orbital polarization overrules the effective molecular fields, which act 
on the $\{a,b\}$ orbitals, already for $D=0.03$ eV (see Fig. 
\ref{fig:11}), and the trend in the ED results changes --- in the regime 
of small $D$, the filling of the occupied $\{1b,2a\}$ orbitals, 
$n_{1b\downarrow}=n_{2a\downarrow}$, first slightly 
increases with increasing $D$ with respect to the one at $D=0$, but next
when $D$ increases further it starts to decrease and the orbital
entanglement is suppressed. In this latter regime, one again finds
a perfect agreement between the HF and the ED results, which shows
once more that the HF method is very reliable for states with broken
symmetry. Note that, as long as the orbital fluctuations dominate for 
very small values of $D<0.02$ eV, the orbital electron densities 
almost do not change, being stabilized by the orbital entangled state
that also counteracts the rotation of $c$ orbitals in this regime.
On the contrary, in the regime of large $D>0.1$ eV, the electron
densities are very similar to those found in Sec. \ref{sec:atom}
for an embedded atom, cf. Fig. \ref{fig:5}.

\begin{figure}[t!]
\includegraphics[width=7.8cm]{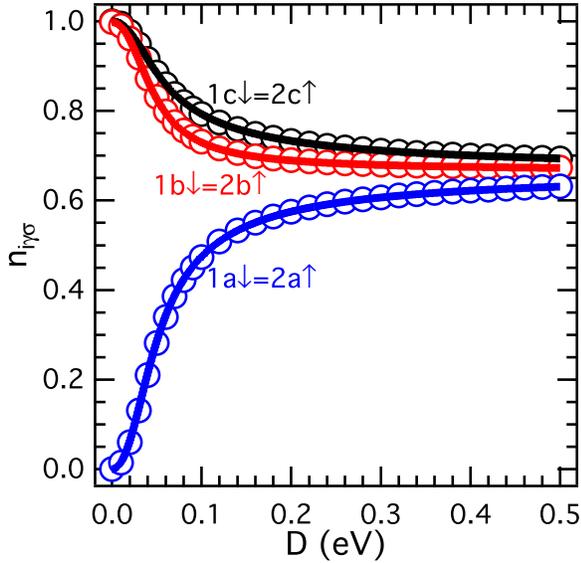}
\caption{(Color online) 
Occupations of pristine orbitals $n_{i\gamma\downarrow}$ at the 
embedded bond $\langle ij\rangle$ along the $c$ axis in the $GC$ 
phase, see Fig. \ref{fig:4}(c), as functions of increasing orbital 
polarization $D$. Solid lines from HF and hollow circles from ED. 
Parameters as in set B in Table \ref{tab:para}.}
\label{fig:12}
\end{figure}

It is of interest now to compare the evolution of electron densities
with increasing $D$ between the $CG$ and the $GC$ phases. The latter
phase features much less quantum fluctuations, as the AF spin order
is present here also along the $c$ axis and finite electron hopping
would generate LS states unfavored by Hund's exchange. Therefore,
the hopping is suppressed also along vertical bonds in agreement with
the double exchange mechanism,\cite{vdB99} at least in the HF picture
where quantum fluctuations are almost completely neglected. Therefore,
even when the electron hopping is finite, the kinetic energy is totally
suppressed in this symmetry-broken state at $D=0$, and the electron
densities are as in Eqs. (\ref{frozen}), see Fig. \ref{fig:12}.
When $D$ increases, the orbital rotation towards the optimal orbital 
basis in the limit of large $D$ takes place, see Fig. \ref{fig:8}, and 
the orbitals are again locally adjusted by the orbital polarization 
term. Notably, the electron densities for
the initially filled $\{1b,2a\}$ and empty $\{1a,2b\}$ orbitals
are here more distinct than in the $CG$ phase as the kinetic energy,
which helps to make the electronic distribution almost symmetric,
is absent. Altogether, the excellent agreement found in this case
between the HF and the ED results confirms the more classical character
of the $GC$ phase with respect to the $CG$ one.

\subsection{One-particle excitations\label{sec:bondex}}
\label{sec:exci}

\begin{figure}[t!]
\includegraphics[width=7.8cm]{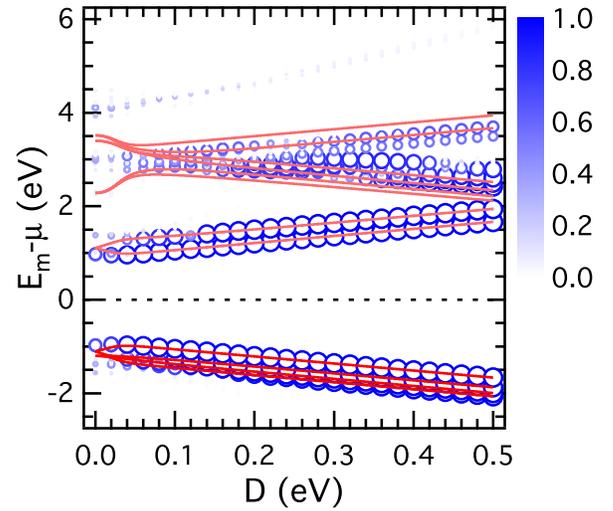}
\caption{(Color online) 
Excitation energies $E_{m}$ for the embedded bond $\langle ij\rangle$ 
along the $c$ axis within the $CG$ phase, see Fig. \ref{fig:4}(b), as 
functions of increasing orbital polarization $D$. Solid lines from HF 
(heavy/light lines for occupied/unoccupied states), 
and hollow circles from ED (the intensity scale and marker size gives 
the spectral weight of each level). 
Parameters as in set B in Table \ref{tab:para}.}
\label{fig:13}
\end{figure}

We determined also the excitation energies, defined in Eqs. 
(\ref{exhole}) and (\ref{exelec}), for the embedded bond. Overall, the 
results found for the $CG$ phase, see Fig. \ref{fig:13}, resemble those 
obtained for an embedded atom, see Fig. \ref{fig:9}. A qualitatively 
new feature in the ED is the broadening of the Hubbard subbands, found 
for the LHB and for the individual excitations that belong to the UHB. 
This broadening comes from the action of the kinetic energy which mixes 
the occupied states at both sites of the bond $\langle ij\rangle$ 
leading to bonding and antibonding states that are modified by the JT 
effective interactions. As for the atom, $U$ is the dominating energy 
scale aand the MH gap increases linearly with $D$. 
There are four excitation energies 
that belong to the LHB and stem from two excitations at each ion of the 
considered bond. Having these splittings in the one-electron states, it 
is natural to expect that also the three-electron states will occur in 
certain energy intervals for each (HS or LS) excited state. The number 
of excited states in the HS states is two as there could be two states
filled by three electrons, one at each site. 

The analysis of the LS states is more subtle here than in the embedded 
atom case, as not only four LS2 states (two at each atom) split off 
with increasing $D$, but because also their spectral weight is 
transferred from the LS2 states to energies characteristic of the 
LS1 excitations. All in all, the qualitative picture found before 
for the embedded atom is confirmed here for the embedded bond, and again
the LS1 excitations, which correspond to creating doubly occupied
orbitals of $b$ character or single occupations in the $a\uparrow$
orbital, stay roughly in the same distance from the LHB independently
of the value of $D$, while the LS excitations at higher energies are
finally the ones that involve double occupancies in the $c$ orbital.
The distance between these states and the HS excitations is close to 
$3J_{H}$, the value deduced from the multiplet splitting.
\cite{Ole05}
We remark that the HF approximation gives excitation energies that
reproduce quite well the values obtained within the ED. This holds
in particular for the LHB and for the HS states of the UHB.

\begin{figure}[t!]
\includegraphics[width=7.8cm]{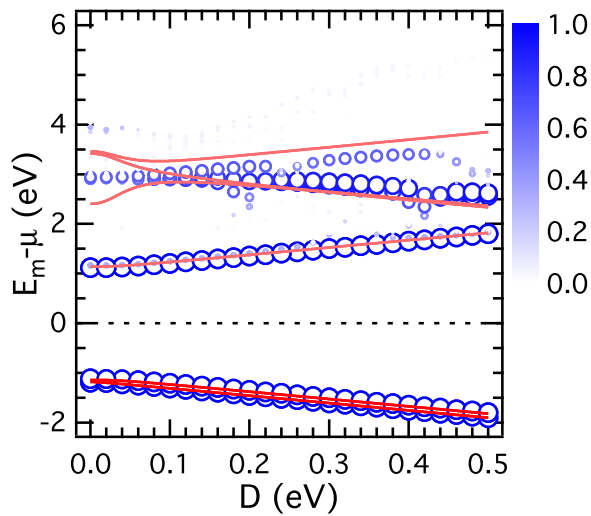}
\caption{(Color online) 
Excitation energies $E_{m}$ for the embedded bond
$\langle ij\rangle$ along the $c$ axis within the $GC$ phase, see
Fig. \ref{fig:4}(c), as functions of increasing orbital polarization
$D$. Solid lines from HF (heavy/light lines for occupied/unoccupied
states), and hollow circles from ED (the intensity scale and marker
size indicate the spectral weight of each level). Parameters as in
set B in Table \ref{tab:para}.}
\label{fig:14}
\end{figure}

Given the more classical character of the $GC$ phase, the LHB and
the UHB are even more pronounced in this case, see Fig. \ref{fig:14}.
Here, the splittings arising from the finite kinetic energy and found
in the $CG$ phase are absent. Accordingly, the dependence of the
one-hole (PES) and the one-electron (IPES) excitation energies is
very similar to those found for a single embedded atom in this phase
(not shown). The LHB consists of two excitations that correspond to
adding a hole to one of the occupied orbitals --- either the $c$
orbital, at the lower energy because of the crystal field, or the
orbital in the $\{a,b\}$ doublet locally favored by the JT
interactions. These excitations have been found both in the ED and
in the HF at exactly the same energies.

The situation for the electron excitations at high energy is again
more subtle. Here one finds in ED four LS excitations at energy $U$
with respect to the ground state, and they split off with increasing
orbital polarization $D$ --- the ones with a roughly constant energy
distance from the LHB do correspond to the double occupancy of $b$
orbital or single occupations in the $a\uparrow$ orbital, while the 
ones that follow from double occupancy of $c$ orbital exhibit a similar 
energy increase with increasing $D$ as the HS state. Already in the
range of relatively small $D\sim0.05$ eV, the spectral weight is
transferred from the LS2 excitations to lower energies, and the 
excitation energy of $U+2J_{H}$ is not observed in the ED results 
anymore.

Here again the HF approximation reproduces all trends found in the
ED results, but instead of the spectral weight transfer of the ED,
one finds that the evolution of the HF wave functions simulates very
well the ED findings at sufficiently large values of $D$. Systematic
errors in the HF, due to the absence of quantum fluctuations, have been
found only in the regime of small $D<0.1$ eV, where the HF energies
interpolate between the excitation energies $U-J_{H}$ and $U+J_{H}$
(lower by $J_{H}$ than the corresponding ED values at $D=0$) and the 
values expected from the ED at sufficiently large $D$. Altogether, the 
agreement between the HF and ED is better in the regime of large $D$, 
where the quantum effects in the orbital space are suppressed.

\subsection{Localization by Coulomb interaction $U$}
\label{sec:bondu}

So far, the results have shown that the HF approximation provides
a realistic description of the electron density distribution in the
occupied orbitals as well as of the excitation energies. This latter
feature is not guaranteed in the HF approach by itself, but follows here
from the relatively large value of $U=4$ eV, which drives the electron
localization and suppresses quantum fluctuations (except for spin
fluctuations, fluctuactions related to the ``pair hopping'' term
in Eq. (\ref{HU}), and the orbital fluctuations along FM bonds). Here
we show that the results obtained for the selected parameters are
actually valid in a broad range of values for the Coulomb parameter $U$.

\begin{figure}[t!]
\includegraphics[width=7.8cm]{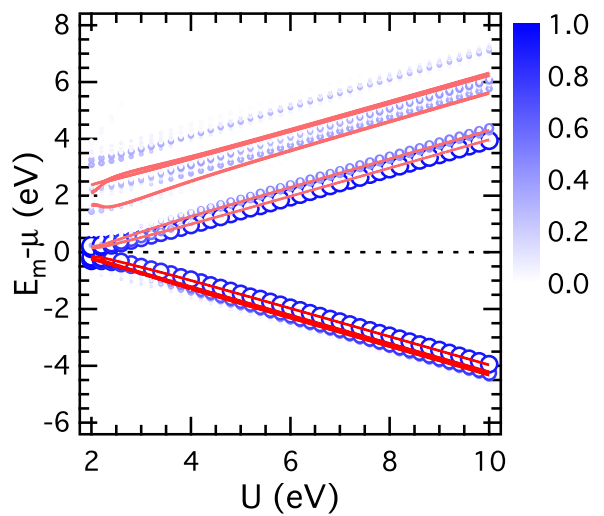}
\caption{(Color online) 
Energy levels in the embedded bond $\langle ij\rangle$
along the $c$ axis within the $CG$ phase, see Fig. \ref{fig:4}(b),
as functions of increasing Coulomb interaction $U$ for $D=30$ meV.
Solid lines from HF and hollow circles from ED 
(the intensity scale and marker size indicate the spectral weight 
of the level). Other parameters as in set B in Table \ref{tab:para}.}
\label{fig:15}
\end{figure}

We fix Hund's exchange parameter $J_{H}=0.6$ eV, which is responsible
for the HS value ($S=1$) at V$^{3+}$ ions in the ground state of the 
undoped system YVO$_{3}$. Given this choice, we varied the parameter $U$ 
in a quite broad range of values starting from $U=2.0$ eV.\cite{noteu} 
As we have shown in Fig. \ref{fig:13}, a relatively small value of 
$D=0.03$
eV suffices to suppress the orbital fluctuations along the considered
$c$ bond in the $CG$ phase. In this regime of parameters, the excitation
energies $E_{m}$ given by the HF approximation are in reasonable
agreement with the ED ones (see Fig. \ref{fig:15}), and only the
LS states have systematically lower energies than their ED counterparts,
as discussed in Sec. \ref{sec:atom}. As a matter of fact, this holds
in the whole range of $U>2.0$ eV, i.e., in the entire physical range
of local interactions relevant for transition metal oxides.\cite{noteu}

\begin{figure}[t!]
\includegraphics[width=7.8cm]{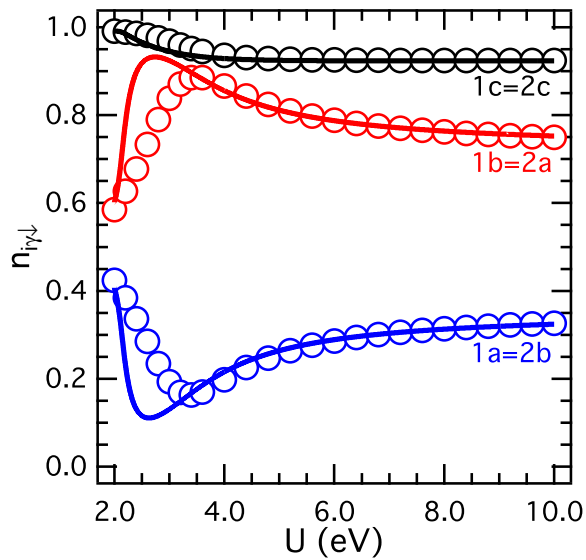}
\caption{(Color online) Occupations of spin-down pristine orbitals 
within the embedded bond $\langle ij\rangle$ along the $c$ axis in the 
$CG$ phase, see Fig. \ref{fig:4}(b), as functions of increasing Coulomb
interaction $U$ for $D=30$ meV. Solid lines from HF and hollow circles
from ED. Other parameters as in set B in Table \ref{tab:para}.}
\label{fig:16}
\end{figure}

The dependences of the one-hole and the one-electron excitations on
$U$ are complemented by the electron densities projected onto the
orbitals of $t_{2g}$ symmetry shown in Fig. \ref{fig:16}. The $c$
orbitals are almost filled in the entire explored range of $U$. They
are relatively robust with respect to orbital polarization for the
present small value of $D=30$ meV, but nevertheless the density $n_{ic}$
gradually decreases from $n_{ic}=1$ at $U=2.0$ eV to $n_{ic}\simeq0.92$
at large $U=10$ eV. This evolution of the electron densities $n_{ic}$
follows from the behavior of the orbital doublets $\{a,b\}$ at both
sites.

First, in the regime of small $U$, electrons are not localized as
the charge excitation energy is the same as the hopping parameter.
The ED gives then the densities 0.58 (0.42) for the more (less) filled
orbitals at the lowest considered value of $U=2.0$ eV. The HF approach 
almost perfectly reproduces these densities. Increasing $U$ suppresses 
the electron hopping along the bond, and the electron densities are  
gradually modified. The many-electron quantum state found in the ED is 
mainly determined by the fields acting on the orbital system: the JT 
effective fields from the neighboring atoms and the orbital polarization 
term. In the regime of $U<3.5$ eV the change of the electron densities 
is markedly slower in the ED than in the HF approximation, as here 
quantum fluctuations contribute.
Only for higher values of $U$, both calculations agree with each other 
showing once more that the HF approximation is reliable in the entire 
strongly-correlated regime of large $U$ due to the stabilization of 
symmetry breaking in spin-orbital space which suppresses quantum
fluctuations.

\subsection{Ground state for the embedded bond}
\label{sec:bondgs}

\begin{figure}[t!]
\includegraphics[width=8.2cm]{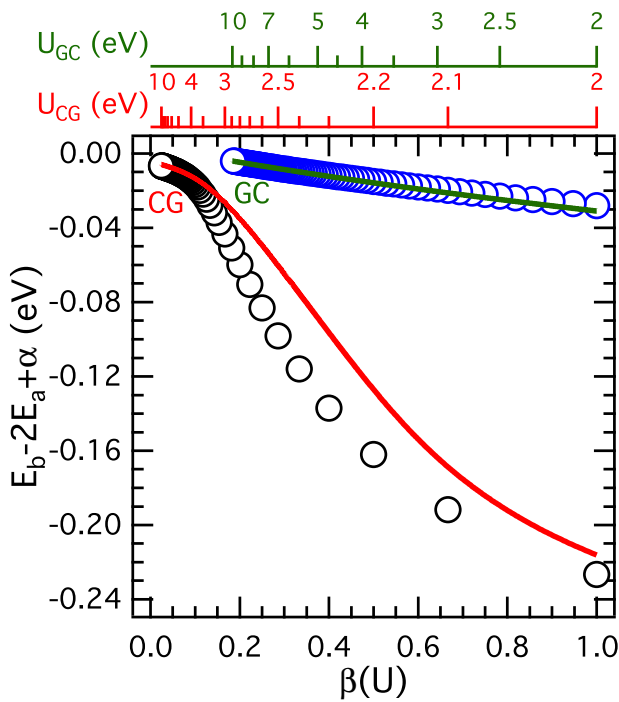}
\caption{(Color online) Hartree-Fock results (solid lines) for the 
superexchange contribution to the binding energy ${\cal E}_{\rm sex}(U)$ 
are compared to ED (circles) for an embedded bond $\langle ij\rangle$ 
along the $c$ axis in the $CG$ and the $GC$ phases for $D=30$ meV. 
For convenience the $U$ dependence is expressed by 
a nonlinear scale (top) and by the perturbative scaling functions  
$\beta_{{\rm CG}}(U)$ (\ref{ecg}) and $\beta_{{\rm GC}}(U)$ (\ref{egc}), 
respectively.
Other parameters as in set B in Table \ref{tab:para}.}
\label{fig:17}
\end{figure}

One may wonder how accurate the HF approximation is for the total
energy of the considered bond. When the bond forms, one gains binding
energy due to the kinetic energy. The binding energy can be computed
as the energy difference between the energy of a bond and that of
two single atoms. 
\begin{equation}
{\cal E}(U)\equiv E_{\rm bond}(U)-2E_{\rm atom}(U).\label{ebin}
\end{equation}
Thereby, one has to take into account the different numbers of neighbors 
in each case. An atom is surrounded by six neighbors, each of them 
providing mean fields acting on the orbitals. In contrast, each atom of 
the bond has just five external neighbors, while the interactions along 
the bond are rigorously (approximately) included in the ED (HF). This 
generates a correction term $\alpha$ to Eq. (\ref{ebin}) included in 
Fig. \ref{fig:17}, where we analyze the superexchange contribution 
to the binding energy ${\cal E}_{ex}(U)$ defined as
\begin{equation}
{\cal E}_{\rm sex}(U)\equiv {\cal E}(U)-{\cal E}(\infty).\label{esex}
\end{equation}

The binding energy ${\cal E}(U)$ (\ref{ebin}) calculated in 
the ED can be compared with the strong-coupling approach of Ref. 
\onlinecite{Ole07} to estimate the energy increments for the embedded 
bond and for the embedded atom in both magnetic phases, the $CG$ and 
the $GC$ phases. In the regime of large $U\gg t$, one finds asymptotic 
behaviors that suggest to use:
\begin{eqnarray}
\beta_{{\rm CG}}(U) & \equiv & \frac{t}{U-3J_{H}}\,,\label{ecg}\\
\beta_{{\rm GC}}(U) & \equiv & \frac{9t}{U-\tfrac{1}{3}J_{H}}\,,\label{egc}
\end{eqnarray}
as scaling parameters. The value of $\beta_{{\rm CG}}(U)$ (\ref{ecg})
is obtained by considering HS excitations, while the value of 
$\beta_{{\rm GC}}(U)$ (\ref{egc}) is derived as an average value over 
the LS excitations allowed in this case. These quantities have been used 
to plot the superexchange energy ${\cal E}_{\rm sex}(U)$ (\ref{esex}) 
for a single bond embedded in each magnetic phase considered here, see 
Fig. \ref{fig:17}. 
We show the data for the same range of $U$ values ($2$ eV $\leq U\leq10$ 
eV) adopted in Figs. \ref{fig:15} and \ref{fig:16}, respectively, which 
corresponds to $\beta(U)\leq1.0$ in both phases, and 
 $\beta_{{\rm CG}}(U)>0.024$ 
($\beta_{{\rm GC}}(U)>0.183$) in the $CG$ ($GC$) phase.

The analysis of the electron density distribution in Sec. 
\ref{sec:bondu} suggests that the local Coulomb interactions are 
sufficiently strong to localize the electrons if $U>3.5$ eV, see Fig. 
\ref{fig:16}, and one expects that the strong-coupling expansion in 
powers of $t/U$ should be valid. This is also confirmed by the present 
analysis of the binding energy ${\cal E}(U)$ (\ref{ebin}) --- the 
numerical results obtained from the ED and the HF method agree very well 
in this regime for the $CG$ phase, see Fig. \ref{fig:17}. Even more 
surprising is the fair agreement between the ED and the HF calculations 
found for smaller values of $U<3.5$ eV, see Fig. \ref{fig:17}. 
This result shows 
that the strong orbital quantum fluctuations in the $CG$ phase along the 
$c$ axis \cite{Kha01} are well accounted for within the unrestricted HF 
approach. In the $GC$ phase, the agreement is even better and the 
dependence on $\beta_{{\rm GC}}(U)$ is practically linear in the entire 
regime of considered values of $U$, i.e., for $U>2.0$ eV. This confirms 
a quite strong electron localization in this phase, with the AF spin 
order preventing electronic transport and acting as a confining 
potential at each site of the bond.

\section{Defects in the dilute limit}
\label{sec:dop}

\subsection{Orbital polarization and the electronic structure}
\label{sec:dopd}

The role of defects in transition metal oxides is subtle due to strong
electron correlations, which manifest themselves in the multiplet 
structure of the MH bands. As a result of the orbital degeneracy, strong 
orbital polarization and relevant relaxation processes occur in the 
vicinity of charged defects. Since a Ca-defect replaces an Y-ion in 
YVO$_3$, each defect has eight vanadium neighbors that form a cube shown 
in Fig. \ref{fig:1}; hence, the orbital degeneracy, which is controlled 
by the spin-orbital order in these compounds, is locally strongly 
affected by the defects.\cite{Hor11} 
The (occupied) orbitals in the neighborhood of 
the defect are polarized, i.e., rotated. This leads to a violation of 
the flavor conservation and, consequently, to a noticeable modification 
of the hopping matrix in the rotated $t_{2g}$ basis. Moreover, for each 
Ca-defect a hole appears in its proximity, bound to it by the Coulomb 
potential of the charged defect itself. Since these holes can move, 
controlled by the double exchange mechanism, they will affect the 
spin-orbital order further.\cite{Hor11} In addition, the chemical 
potential, due to the doped holes, will lie in a defect band inside the 
MH gap. Finally, also these defect states are subject to strong 
electronic correlations and display their own MH physics.

Our main aim here is to investigate whether the MH multiplet splitting
can be reliably obtained within the unrestricted HF framework for the 
3-flavor $t_{2g}$ case, i.e., in presence of defect potentials, various 
orbital relaxation processes, kinetic energy of doped holes and 
electron-electron interactions. 
Here we consider the dilute limit of low electron doping $x\le 0.05$, 
where the defects can be treated as well separated from one another. 
For more clarity, we also limit ourselves to hole doping for only one 
orbital flavor and one spin. Therefore, to demonstrate this, we focus 
below on several rather transparent cases of increasing complexity:
\begin{itemize}
\item[(A)] defect states in a multiband MH insulator in absence of 
orbital polarization and of Coulomb interaction among the electrons;
\item[(B)] role of orbital polarization;
\item[(C)] role of LR electron-electron interaction --- Coulomb gap 
of defect states;
\item[(D)] orbital polarization and relaxation of Coulomb gap of defect 
states.
\end{itemize}
For the sake of clarity and simplicity, we will not consider here the 
effect of LR defect potentials, and the self-consistent screening of LR 
interactions, which results from the doping induced by the impurities. 
As our focus is on the strong-correlation aspects and in order to get 
easy-to-interpret eigenstates and spectral distributions, we shall not 
discuss the effects of disorder here, i.e., the defects will always be 
well annealed and in the dilute limit.

\subsection{Order parameter landscape and defects}
\label{sec:land}

\begin{figure*}
\includegraphics[width=16cm]{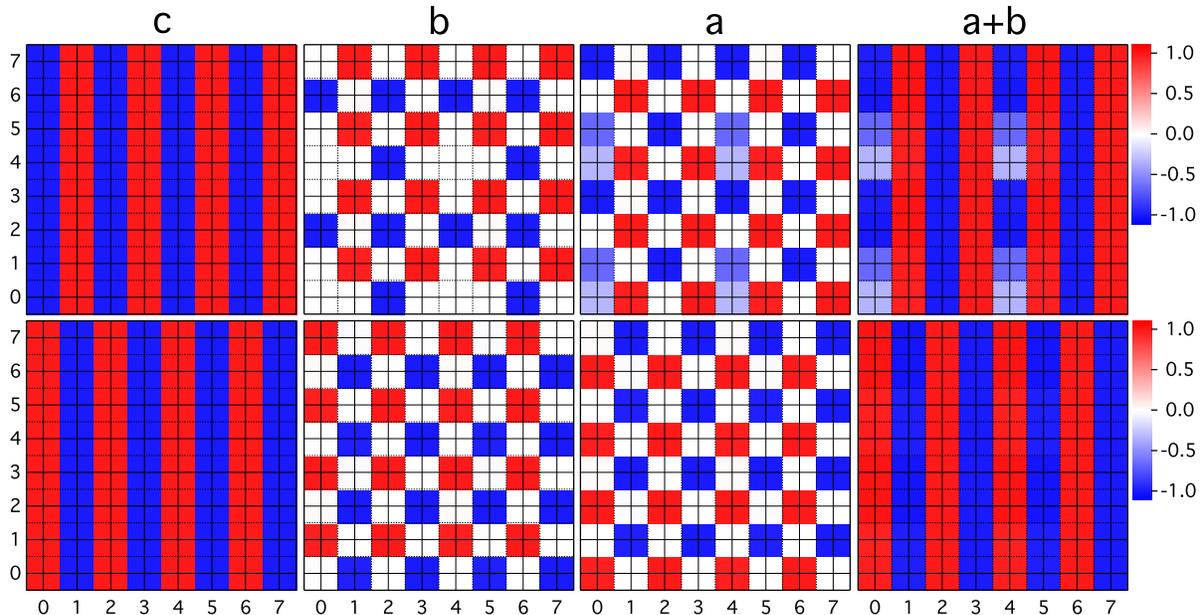}
\caption{(Color online).
Map of the orbital magnetization densities $m_\gamma(k,l,m)$ (\ref{map}) 
in the dilute well annealed $CG$ phase with $x=0.015625=1/64$, obtained 
for a $8\times 8\times 8$ cluster (from positive to negative 
$m_\gamma(k,l,m)$, see color scale on the far right). 
The coordinates 
$k=0,\ldots,7$ are along the horizontal axis, and $m=0,\ldots,7$ along 
vertical axis, while the top row represents the $l=0$ and the bottom 
row the $l=1$ plane. 
Four panels (from left to right) for different orbital flavor: 
$\gamma=c$, $\gamma=b$, $\gamma=a$, and $\gamma=a+b$, respectively.  
Parameters: $V_{\rm ee}=0$, $V_{\rm D}=1$ eV and ${D=0}$, 
others as in set B in Table \ref{tab:para}.}
\label{fig:19}
\end{figure*}

Unrestricted HF approach allows us to treat large inhomogeneous systems, 
i.e., with defects, interfaces, nanostructures, {\it etcetera}. We have 
seen earlier for a two-flavor model that the Mott-Hubbard gap and the 
multiplet structure is faithfully reproduced for the strongly correlated 
transition metal oxide systems. A central requirement of the 
unrestricted HF to work is broken spin and orbital translational 
symmetry. Accordingly, vanadate perovskites, like Y$_{1-x}$Ca$_x$VO$_3$, 
provide an ideal ground for this approach as spin and orbital orders in 
these compounds are present in a wide doping regime, actually up to the 
metal-insulator transition. Therefore, the unrestricted HF is the method 
of choice in our case. Nevertheless, it is not clear {\it a priori} how 
well the method works in the vicinity of defects in view of orbital 
polarization and hole motion. To explore this is our central goal.

Moreover, the HF provides a basis for subsequent many-body perturbation 
treatments. However, it should be noted that here such a
perturbative treatment will not be performed with respect to an 
'uncorrelated' $U=0$ state, but with respect to the HF ground state.
This latter state already features the MH gap and the higher multiplet 
structure. In fact, it was already pointed out that the fundamental MH 
splitting, i.e., the energy difference of the LHB and the high-spin 
Hubbard band $\Delta_{\rm HS}=(U-3J_H)$ is obtained exactly within HF. 
We shall see below that there are other gaps, e.g. between defect states, 
which suffer from the well-known overestimation of gap energies within 
HF-theory. A subsequent many-body perturbation treatments of the 
screening of interactions is expected to cure such deficiences of the 
HF approach.

In general, HF calculations require a self-consistent optimization of 
all expectation values that define the final HF Hamiltonian, the HF 
ground state and the excitations. Complete information to construct the 
HF Hamiltonian is thus contained in the {\it order parameter landscape},
which is defined by the full set of effective fields. It contains 
already the rotations of occupied (and unoccupied) orbitals due to 
defects and doping. As an example, we display the diagonal occupation 
numbers in the $CG$-phase for the 3-flavor calculation performed with
PBC in Fig. \ref{fig:19}. It shows maps of local magnetizations, 
\begin{equation}
m_\gamma(k,l,m)=n_{\gamma\uparrow}(k,l,m)-n_{\gamma\downarrow}(k,l,m),
\label{map}
\end{equation}
for an $8 \times 8 \times 8$ cube of V ions and --- from left to right 
--- for the orbital character $\gamma=c,b,a$ and $a+b$. Two planes are
presented: the top row represents the front face of the cube at $l=0$, 
while the bottom row shows the $l=1$-plane.

The left panels reflect the $C$-type magnetic order of the electrons in 
$c$ orbitals. The central panels show the same magnetic order for the 
$b$ and $a$ electrons. Here, in addition, one recognizes a checkerboard 
structure, which represents the $G$-type orbital order of the $a$ and 
$b$ electrons. In the right panel, magnetic densities of electrons in 
$a$ and $b$ type orbitals are summed up and reveal again the $C$-type 
magnetic structure.
While the magnetization density has a perfect alternation between two 
sublattices in the $l=1$ plane, four defects disturb it in the $l=0$ 
layer (top row), as seen for $b$ and $a$ orbitals, and also for their 
sum. 

The number of defects in the system corresponds in this case to hole
doping $x=1/64$. Here, the arrangement of the defects is well annealed, 
i.e., LR defect repulsion has been considered and, therefore, the 
defects are regularly spaced. The defects are visible in the top panel 
of Fig. \ref{fig:19} and are located between the planes $l=0$ and $l=1$.
The perpendicular feature in the $a+b$-panel reflects the motion of the 
doped holes along the c direction, but bound to the respective defects.
From the $b$-panel, one can also see that all 4 holes in the $l=0$ plane 
have $b$-down character, as assumed. We next analyze the one-particle 
excitation spectra for such a cluster obtained in the HF approach.

\subsection{Excited states in the 3-flavor model}
\label{sec:three}

In the following, we shall investigate the HF excitation spectra for the 
four scenarios (A)-(D) introduced above. We will analyze only the $CG$ 
phase as this phase is the relevant one for moderately doped 
Y$_{1-x}$Ca$_x$VO$_3$, and we first focus on the dilute doping case of
$x=1/64$.

(A) We begin the discussion with the first scenario, where orbital 
polarization effects and LR Coulomb interactions are not considered, 
i.e., $D=0$ and $V_{\rm ee}=0$. The strong defect potential, $V_D=1.0$ 
eV, confines a single doped hole per defect to the nearest neighbor 
V-sites of a defect, which form a cube. The corners of the V$_8$ cube 
are equivalent with respect to the defect potential yet not with respect 
to the spin- and orbital- order. This implies that a $(b\downarrow)$ hole 
can delocalize only along a (vertical) bond along the $c$ axis because 
of the AF order in the (horizontal) $ab$-plane. The kinetic energy of 
the $a$ and $b$ 
electrons in the $ab$-plane is quenched due to the double-exchange-type 
coupling to the spins of the $c$ electrons.\cite{Hor11}

\begin{figure}[tbp]
\includegraphics[width=7cm]{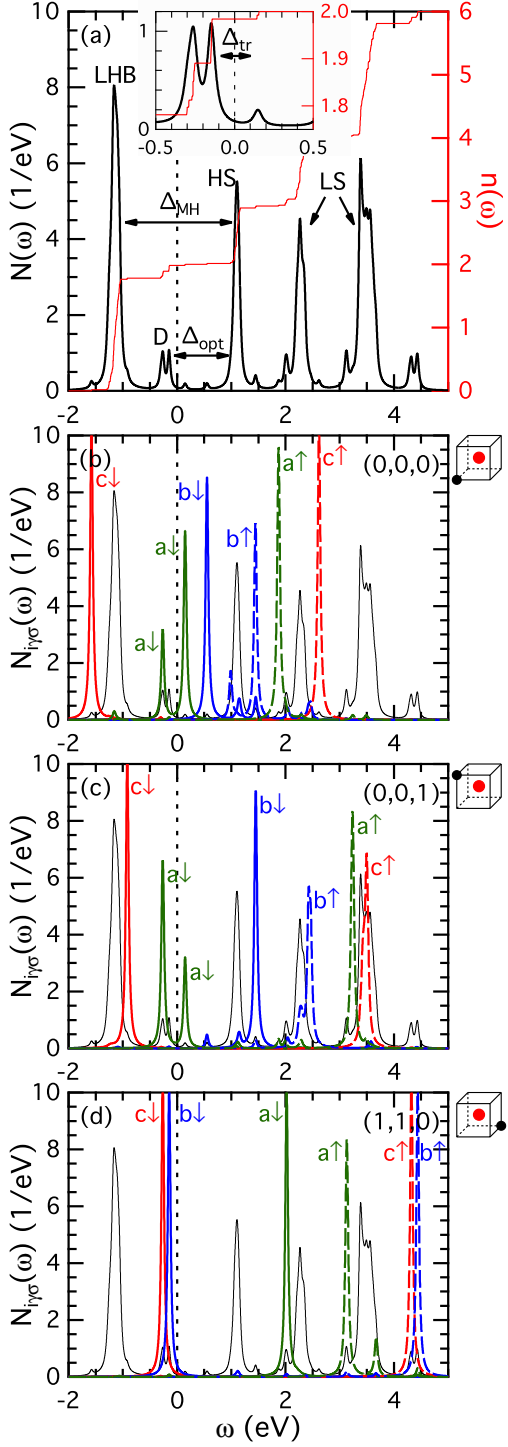}
\caption{(Color online)
(a) Total DOS $N(\omega)$ (left scale) and average electron filling 
$n(\omega)$ (right scale) in 
the dilute well annealed $CG$ phase with defect concentration  
$x=1/64$, obtained using an $8\times 8\times 8$ cluster. The inset shows 
$N(\omega)$ near the Fermi energy $\omega=0$. Panels (b)-(d) show
individual DOSs $N_{i\alpha\sigma}(\omega)$ for orbital $\alpha$ and 
spin $\sigma$, together with the total DOS $N(\omega)$ (thin lines).
The atom coordinate, (0,0,0), (0,0,1), or (1,1,0), is indicated in 
each case and on the cube represented near each panel on the right.
Parameters as in set B in Table \ref{tab:para}, 
and $D=0$, $V_{\rm ee}=0$.}
\label{fig:20}
\end{figure}

To explore this symmetry breaking and its consequences for the 
excitations, we investigate in Fig. \ref{fig:20} the total DOS 
$N(\omega)$ together with the local partial DOS 
$N_{i\gamma\sigma}(\omega)$ for the two sites (0,0,0) and (0,0,1) of 
the {\it active bond}, i.e., the bond doped with a hole, and for a site 
(1,1,0) on an undoped {\it spectator bond}. The chosen spectator bond 
corresponds to the same spin polarization as the active bond, thus in 
principle the hole could delocalize into its states too. However, this 
is not observed for the typical range of values of $t$ (and other 
parameters). The total DOS in Fig. \ref{fig:20}(a), obtained for the 
$8\times 8\times 8$ cluster, shows the same multiplet splitting as 
discussed for the atom in Sec. \ref{sec:atom}. We recognize the LHB and 
the three multiplet subbands forming the UHB. The fundamental MH gap 
opens between the LHB and the HS subband of the UHB; the splitting of 
these bands is given by the gap,
\begin{equation}
\label{HS}
\Delta_{\rm HS}=E_{\rm HS}-E_{\rm LHB}=U-3J_H, 
\end{equation}
and is reproduced exactly in the HF approach. In contrast, the actual 
MH gap,
\begin{equation}
\Delta_{\rm MH}=U-3J_H-W_{\rm eff},
\end{equation}
is reduced by an effective bandwidth $W_{\rm eff}$. Due to the defects 
and their potentials, defect states are split off from the LHB. In other 
words, all single particle states in the vicinity of the defect are 
shifted upwards by the defect potential $V_D$ (for simplicity, we 
consider here ony the case of a nearest neighbor defect potential). 
These states lie now inside of the MH gap. The transitions from the 
occupied defect states $D$ to the upper HS Hubbard subband define the 
ingap absorption and the optical absorption gap which we denote 
$\Delta_{\rm opt}$. 
At the same time, the spectral weight moves out of the LHB, similar as 
in the doped Hubbard model.\cite{Esk94}

The defect states would be completely filled, if there were not one 
doped hole per defect. This fixes the chemical potential $\mu$ inside 
the defect band. It is useful to consider the integrated (averaged) 
electron density $n(\omega)$ per site, Eq. (\ref{nw}), displayed in Fig. 
\ref{fig:20}(a), see right scale. The chemical potential $\mu$ is then 
determined from the average of $\mu^-$ and $\mu^+$ which are obtained 
from the relation,
\begin{equation}
n(\mu^{\pm})= n_0 - x \pm \epsilon,
\end{equation}
in the limit $\epsilon\rightarrow 0$, where $n_0=2$ is the average 
number of electrons per site in the undoped 3-flavor model and $x$ is 
the defect (hole) concentration. As can be seen from the inset in Fig. 
\ref{fig:20}(a), the chemical potential falls into a small gap at the 
upper edge of the defect band $D$. In the following, we denote this gap 
as the {\it transport gap} $\Delta_{\rm tr}$ as it would be a relevant 
measure for the conduction in the defect band.

Deeper insight into the nature of the defect states and the transport 
gap is obtained by considering the local partial DOS 
$N_{i\gamma\sigma}(\omega)$, shown in Figs. 
\ref{fig:20}(b)-\ref{fig:20}(d) for sites (0,0,0) and (0,0,1), which 
form the {\it active bond}, and at the (1,1,0) V ion, a {\it spectator 
site} with two electrons. The defect, indicated by a red dot, lies in 
the center of the cube at $(\frac12,\frac12,\frac12)$. The two electrons 
at the spectator site (1,1,0) occupy the $(c\downarrow)$ and 
$(b\downarrow)$ 
local orbitals as can be seen from Fig. \ref{fig:20}(d). They lie below 
the chemical potential $\mu$ indicated by the perpendicular dashed line.

The doped hole has $(b\downarrow)$ character and is localized on 
the active bond displayed in Fig. \ref{fig:20}(b) and (c). On both sites 
the $(c\downarrow)$ orbitals are occupied, while the remaining 
$(a\downarrow)$ electron can delocalize on the vertical 
bond, i.e., parallel to the $c$ axis, consistently with the FM 
correlations in the $C$-AF phase. This delocalization leads to a 
splitting into bonding and antibonding states on the active bond,
\begin{equation}
\Delta_{\rm BA} \sim 2\sqrt{ V^2_{\rm JT}+t^2},
\end{equation}
where the bonding-antibonding gap $\Delta_{\rm BA}$ is determined by the 
resonance integral $t$ and the JT-interaction
$V_{\rm JT}\approx V_{ab}-\frac{1}{4} V_{c}$ in the case of the $G$-AO 
order. In Figs. \ref{fig:20}(b) and \ref{fig:20}(c), 
$\Delta_{\rm BA}\approx 0.4$ eV is essentially determined by the hopping 
matrix element $t=0.2$ eV while $V_{\rm JT}\approx 0.02$ eV. That the 
$|a\downarrow\rangle$ occupation is asymmetric along the active bond is 
a result of the JT interactions and the underlying $G$-type orbital order.

\begin{figure}[tbp]
\includegraphics[width=7cm]{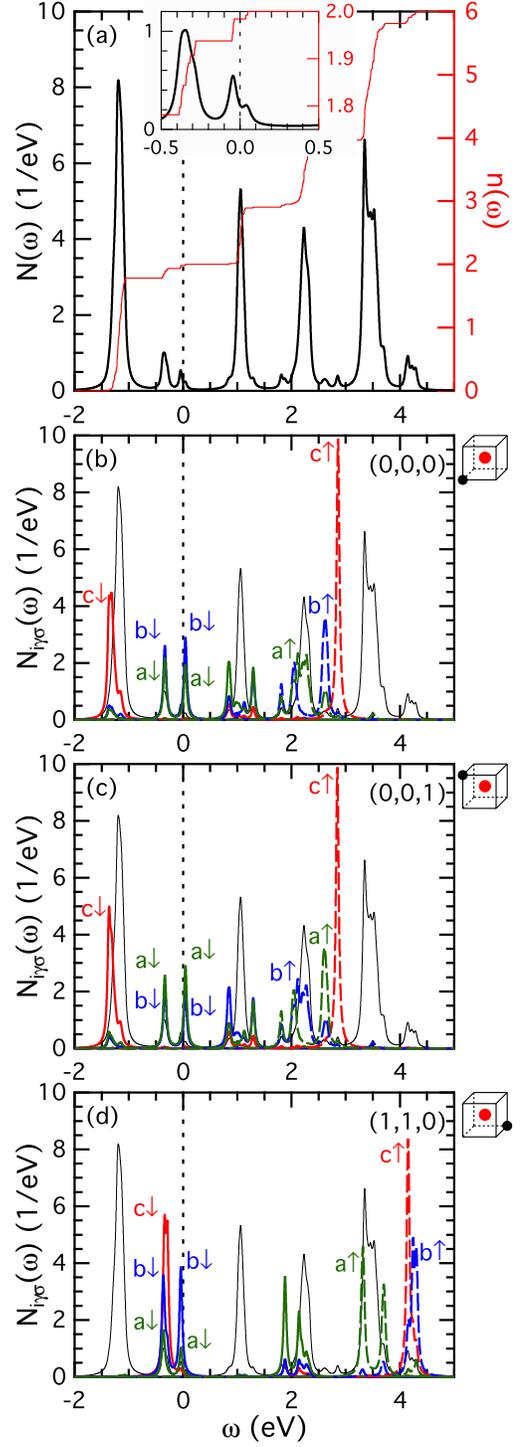}
\caption{(Color online) (a)
Total DOS $N(\omega)$ (left scale) and average electron filling 
$n(\omega)$ (right scale) in 
the dilute well annealed $CG$ phase with defect concentration 
$x=1/64$, obtained using an $8\times 8\times 8$ cluster. The inset shows 
$N(\omega)$ near the Fermi energy $\omega=0$. Panels (b)-(d) show
individual DOSs $N_{i\alpha\sigma}(\omega)$ for orbital $\alpha$ and 
spin $\sigma$, together with the total DOS $N(\omega)$ (thin lines).
The atom coordinate, (0,0,0), (0,0,1), or (1,1,0), is indicated in 
each case and on the cube represented near each panel on the right.
Parameters as in set B in Table \ref{tab:para}, 
and $D=0.05$ eV, $V_{\rm ee}=0$.}
\label{fig:21}
\end{figure}

The origin of the `transport gap' $\Delta_{\rm tr}$ is evident in this 
case and we identify it here with the gap between the topmost occupied 
orbital on a spectator bond, i.e., $|b\downarrow\rangle$ orbital at site 
(1,1,0) and the unoccupied antibonding state on the active bond. This 
energy is
\begin{equation}
\Delta_{\rm tr}\sim \sqrt{ V^2_{\rm JT}+t^2}+V_{\rm JT},
\end{equation}
and for the present parameters one finds in Fig. \ref{fig:20} that 
$\Delta_{\rm tr}\approx 0.22$ eV, and is essentially determined by the 
kinetic energy. It is important to realize that within the HF-scheme 
occupied (unoccupied) orbitals correspond to electron removal (addition) 
energies. These processes need not to refer necessarily to the same 
defect, but can occur at different defects. Thus $\Delta_{\rm tr}$ is 
the energy gap for the hopping processes between different defects 
within the defect band.

The energies of the occupied $(c\downarrow)$ orbitals differ 
strongly at different ions, see Figs. \ref{fig:20}(b)-\ref{fig:20}(d).
This is a consequence of the different local occupation numbers,
which enter the HF-energies. Obviously, this also affects the energies 
of the unoccupied orbitals. The defect states have their own MH gap as 
can be seen for example in Fig. \ref{fig:20}(d). Inserting an 
$(a\downarrow)$ electron on site (1,1,0), in addition to the already 
present $(c\downarrow)$ and $(b\downarrow)$ electrons, leads to the HS 
defect state 
$|c\downarrow b\downarrow a\downarrow\rangle$ at $\omega\sim 2.0$ eV. 
Finally, at higher energies, one recognizes the LS defect multiplet 
states $|c\downarrow b\downarrow \gamma\uparrow\rangle$, 
with $\gamma=a,b,c$, respectively.

(B) The effect of the Coulomb potential of the defect is to rotate the 
occupied orbitals in the vicinity of the defect such that their energy 
is minimized. This polarization effect, described by $H_{\rm pol}$ and 
controlled by the parameter $D$, was not considered in the first 
scenario. The second scenario represents a calculation within the same 
parameter set of the first scenario, with the exception for finite 
$D=0.05$ eV. The total DOS $N(\omega)$ shows a clear change of the DOS 
of the defect states in the vicinity of $\mu$, which is amplified in the 
inset of Fig. \ref{fig:21}(a). While there was a pronounced transport 
gap in the absence of orbital polarization, here this gap has almost 
disappeared (or is of marginal size).

The main effects of orbital polarization can be revealed by inspecting
the local partial DOS $N_{i\gamma\sigma}(\omega)$. As we observe in 
Figs. \ref{fig:21}(b) and \ref{fig:21}(c), which reflect the changes on 
the sites of the active bond, the $(c'\downarrow)$ orbital has now 
some admixtures of $(a\downarrow)$ and $(b\downarrow)$ flavors. More 
importantly, Figs. \ref{fig:21}(b) and \ref{fig:21}(c) show that the 
bonding and antibonding states are defined by linear combinations of 
$|a\downarrow\rangle$ and $|b\downarrow\rangle$ states, and have no 
longer pure $(a\downarrow)$ character, as it was the case in Figs. 
\ref{fig:20}(b) and \ref{fig:20}(c). This mixing is a 
consequence of the flavor non-conservation in the rotated $t_{2g}$ basis 
and of the resulting off-diagonal terms in the kinetic energy.

The most important feature in this second (B) scenario is the 
formation of a bonding-antibonding splitting on the {\it spectator bond}, 
as can be seen in Fig. \ref{fig:21}(d). At first glance, this may look 
puzzling as both states are occupied, so there is no obvious net energy 
gain that may drive such a splitting. Inspection of energies and 
occupation numbers shows that the level-splitting is due to a combined 
mechanism: the rotation of the $c$ orbital is favored when its energy 
coincides with the level of the bonding orbital. Thus, we observe here 
an {\it orbital rotation induced by bonding-antibonding splitting\/} 
on the spectator bonds.

Interestingly, the antibonding band on the spectator bond is pinned to 
(but remains below) the chemical potential, whereas the antibonding 
state on the active bond is confined at (but remains above) the chemical 
potential. This leads to a vanishing transport gap $\Delta_{\rm tr}$. 
However, this 
also implies a remarkable constraint in the system; namely, the hole 
cannot delocalize from the active bond into the opposite spectator bond 
with the same spin orientation.

Another quite important consequence of the orbital flavor mixing is the 
collapse of the charge polarization on the active bond. Whereas in the 
absence of orbital polarization in scenario (A) there is a pronounced
accumulation of $a$-type charge at site (0,0,1) and of hole-density at 
site (0,0,0) triggered by the JT interaction, one observes an 
essentially complete collapse of the combined $a$ plus $b$ charge 
polarization on the active bond in the case of $D=0.05$ eV, see Figs. 
\ref{fig:21}(b) and \ref{fig:21}(c).

\subsection{Role of long-range Coulomb interaction}
\label{sec:dopv}

Next we turn to the effects of the LR Coulomb interaction in
presence of defect states.
In insulators the Coulomb interaction is screened by the dielectric 
constant $\varepsilon_0$ of the system; yet, the interaction keeps its 
$\sim 1/\varepsilon_0 r$ LR character. Here $\varepsilon_0$ 
represents the background dielectric screening due to the 
'core'-electrons. These are in our case all electrons apart from those 
of $t_{2g}$-type. The screening arising from $t_{2g}$ electrons, which 
is for instance essential for the MIT, is 
explicitly included in the Hamiltonian of the system.

\begin{figure}[t!]
\begin{center}
\includegraphics[width=7.8cm]{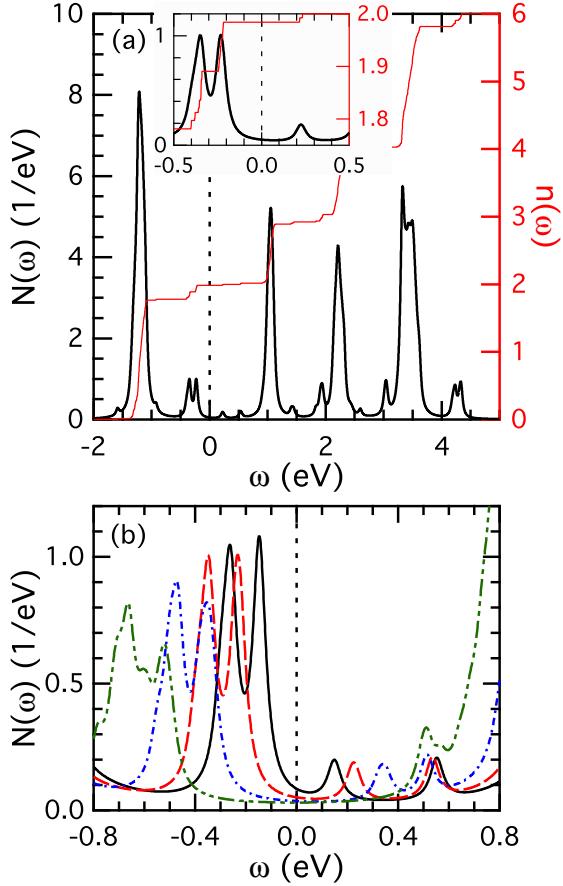}
\end{center}
\caption{(Color online) (a) Total DOS $N(\omega)$ (left scale) and 
average electron filling $n(\omega)$ (right scale) in the dilute well 
annealed $CG$ phase at hole doping $x=1/64$, obtained using an 
$8\times 8\times 8$ cluster.
The inset shows $N(\omega)$ near the Fermi energy $\omega=0$. 
Calculation includes the LR Coulomb interaction $V_{\rm ee}=0.2$ eV.
(b) Dependence of DOS $N(\omega)$ of defect states and Coulomb gap near 
the Fermi energy $\omega=0$ on increasing strength of LR Coulomb 
interaction: $V_{\rm ee}=0$ (solid, black), $0.2$ eV (dashed, red), 
$0.5$ eV (dashed-dotted, blue), $1.0$ eV (dashed-double-dotted, dark green).
For increasing strength of $V_{\rm ee}$ the features in $N(\omega)$ move 
away from the Fermi energy $\omega=0$.
Other parameters as in set B in Table I and $D=0$.}
\label{fig:22}
\end{figure}

In the next scenario (C), the LR Coulomb interaction Eq. (\ref{V1}) is 
taken into account and parametrized via $V_{\rm ee}$ (instead of 
$\varepsilon_0$), while the orbital polarization is neglected ($D=0$), 
as in scenario (A). The DOS is displayed in Figs. \ref{fig:22}(a) and 
\ref{fig:22}(b). Its comparison with Fig. \ref{fig:20}(a) shows that the 
LR Coulomb interaction has no significant effect on the ionic multiplet 
structure, as one might expect. There is, however, a significant change 
in the size of the energetic splittings between the different defect 
states in the vicinity of the chemical potential, as one can see from 
the insets in Figs. \ref{fig:20}(a) and \ref{fig:22}(a). The fine 
structure of the levels is here basically as in Figs. 
\ref{fig:20}(c)-\ref{fig:20}(d). The peaks in the inset stem from the 
occupied $\{(c\downarrow),(b\downarrow)\}$ orbitals at the spectator 
sites as well as from the $(a\downarrow)$ bonding/antibonding 
levels at the active bond. The main effects of the nonlocal LR 
electron-electron Coulomb interaction are: 
(i) to enlarge the bonding-antibonding splitting, and 
(ii) to increase the polarity of the active bond.

\begin{figure}[t!]
\begin{center}
\includegraphics[width=8cm]{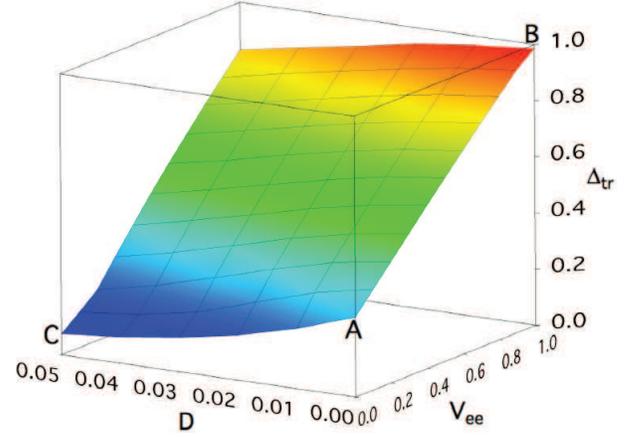}
\end{center}
\caption{(Color online). Transport gap $\Delta_{\rm tr}$ in the 
well annealed $CG$-AO phase with doping concentration
$x=1/64$ as a function of electron-electron interaction coupling
$V_{\rm ee}$ and orbital polarization strength $D$. 
Results obtained for an $8 \times 8 \times 8$ cluster
with the parameters as in set B of Table I. }
\label{fig:24}
\end{figure}

The transport gap $\Delta_{\rm tr}$ opens between the topmost occupied 
$|c\downarrow b\downarrow\rangle$ states on the spectator site and the 
unoccupied $|a\downarrow\rangle$ antibonding state on the active bond. 
Figure \ref{fig:22}(b) shows that it increases almost linearly with 
increasing LR electron-electron interaction $0<V_{\rm ee}<1.0$ eV. 
Indeed, this dependence of $\Delta_{\rm tr}$ on $V_{\rm ee}$ is 
summarized in Fig. 
\ref{fig:24}. The line $A$-$B$ shows the evolution of $\Delta_{\rm tr}$ 
with the increasing value of $V_{\rm ee}$ for vanishing polarization 
field $D=0$. In the absence of nonlocal electron-electron Coulomb 
interaction, the origin of the gap is the bonding-antibonding splitting 
due to the kinetic energy gain of the single $a$-electron on the active 
bond. With increasing $V_{\rm ee}$, the charge polarization of the bond 
grows and the kinetic energy is suppressed. At large $V_{\rm ee}$, the 
gap $\Delta_{\rm tr}$ is determined by the nonlocal Coulomb interactions, 
i.e., essentially by the nearest- and further neighbor ones. The 
increase of $\Delta_{\rm tr}$ along the line $A$-$B$ in Fig. \ref{fig:24} 
may be approximated in terms of the electron-electron interaction 
parameter $V_{\rm ee}$, the hopping integral $t$ and the JT coupling 
$V_{\rm JT}$ as follows,
\begin{equation}
\label{tr}
\Delta_{\rm tr}\approx 
\sqrt{ (V_{\rm ee}/2+V_{\rm JT})^2+t^2}+ V_{\rm ee}/2+V_{\rm JT}.
\end{equation}

We emphasize that, in these systems, the Coulomb gap in the defect 
states is a consequence of the complex structure of the defects in 
combination with the electron-electron interactions. We note that this 
mechanism is distinct from the disorder-induced Coulomb gap of 
Shklovskii and Efros.\cite{Efr75} We have already seen that the doped 
holes in the dilute doping regime have no effect on the multiplet 
splitting away from the defects. On the other hand, on general grounds 
we would expect that the present Coulomb gap in the defect band is 
overestimated within the HF approximation, and will be reduced by 
dielectric screening, i.e., obtained by a many-body treatment beyond the 
HF approach. We shall see next that already the inclusion of orbital 
polarization gives rise to a substantial screening and reduction of the 
Coulomb gap.

\begin{figure}[t!]
\begin{center}
\includegraphics[width=7.8cm]{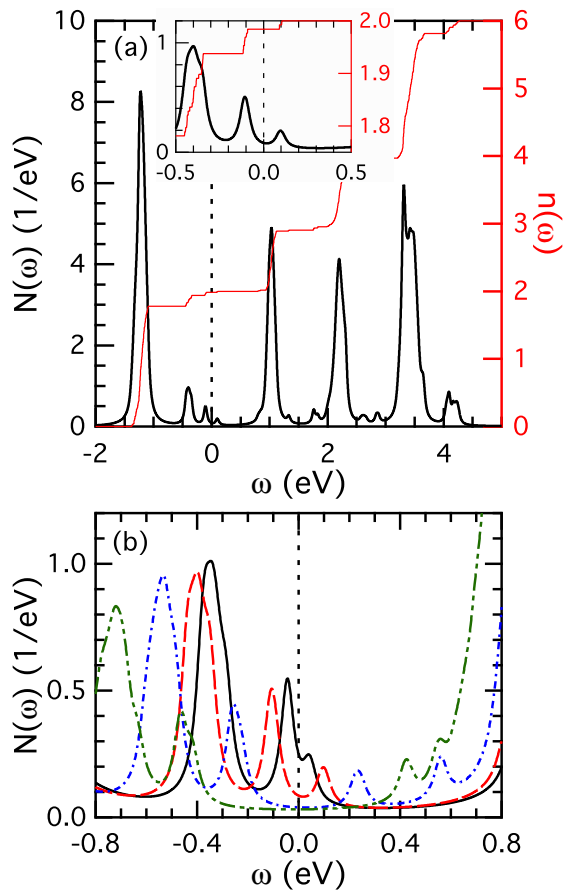}
\end{center}
\caption{(Color online) (a) Total DOS $N(\omega)$ (left scale) and 
average electron filling $n(\omega)$ (right scale) in the dilute well 
annealed $CG$ phase with $x=1/64$, as obtained using an 
$8\times 8\times 8$ cluster for finite LR  Coulomb interaction 
$V_{\rm ee}=0.2$ eV. 
The inset shows $N(\omega)$ near the Fermi energy $\omega=0$.
(b) DOS $N(\omega)$ of defect states and Coulomb gap near the Fermi 
energy $\omega=0$ for increasing strength of LR Coulomb interaction: 
$V_{\rm ee}=0$ (solid, black), $0.2$ eV (dashed, red), $0.5$ eV 
(dashed-dotted, blue), $1.0$ eV (dashed-double-dotted, dark green).
For increasing strength of $V_{\rm ee}$ the features in $N(\omega)$ move 
away from the Fermi energy $\omega=0$.
Other parameters as in set B in Table I and $D=0.05$ eV.}
\label{fig:23}
\end{figure}

Finally, we consider scenario (D) that includes both LR 
electron-electron Coulomb interaction and orbital polarization effects 
in the vicinity of the defects. The DOS is displayed in Figs. 
\ref{fig:23}(a) and \ref{fig:23}(b). Comparison of the insets in Figs. 
\ref{fig:23}(a) and \ref{fig:22}(a) shows the strong reduction of 
$\Delta_{\rm tr}$ in the present case of $D=0.05$ eV from the value 
found with $D=0$. Similarly, Figs. \ref{fig:23}(b) and \ref{fig:22}(b) 
make apparent the strong reduction of the gap at $D=0.05$ eV not only 
for $V_{\rm ee}=0$, but also for larger values of $V_{\rm ee}$.

To summarize, the central observations concerning the transport gap are:
(i) the decrease of $\Delta_{\rm tr}$ with increasing polarization $D$, 
and
(ii) the increase of $\Delta_{\rm tr}$ with increasing LR 
electron-electron Coulomb interaction strength $V_{\rm ee}$.
These findings are summarized in Fig. \ref{fig:24}. 
The strong impact of the orbital polarization $D$-term on 
$\Delta_{\rm tr}$ is seen between the $A$ and $C$ points, where the 
transport gap is reduced by a factor $\sim 3$ when $D$ increases from 
0 to 0.05 eV. This strong suppression of the transport gap change found 
here for relatively weak polarization interaction $D$ is consistent 
with fast orbital rotation induced by this term, presented in Sec. 
\ref{sec:atom}.

\section{Discussion}
\label{sec:dis}

\subsection{Gaps in the electronic structure}
\label{sec:gap}

\begin{figure}[t!]
\begin{center}
\includegraphics[width=7.8cm]{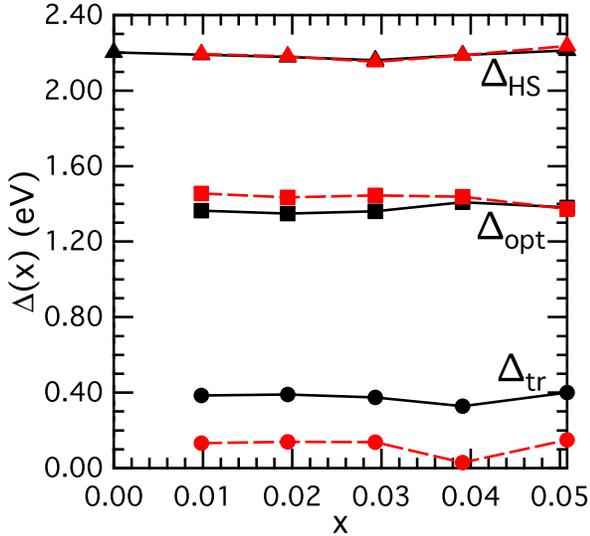}
\end{center}
\caption{(Color online)
Three characteristic gaps in the spectra for increasing doping $x$:
the Mott gap $\Delta_{\rm HS}$ Eq. (\ref{HS}) deduced from the energy 
of the HS transitions (triangles), the optical gap $\Delta_{\rm opt}$ 
between the defect states and the HS UHB (squares), and the transport 
gap $\Delta_{\rm tr}$ within the defect states (circles), determined 
from the flat region in $n(\omega)$ at $\omega=0$.
The three gaps are obtained for the dilute well-annealed doped $CG$ 
phase using an $8\times 8\times 8$ cluster, and for two values of $D$:
$D=0$ (solid lines, black) and $D=0.05$ eV (dashed lines, red).
Parameters: as in set B in Table I and $V_{\rm ee}=0.2$ eV.}
\label{fig:25}
\end{figure}

In our HF study we have encountered three fundamental energy scales 
typical of doped MH insulators. They are displayed in Fig. \ref{fig:25} 
for the dilute doping regime $x\le 0.05$ investigated here. In 
decreasing-size order, these energies represent: 
(i) the Mott gap computed as the energy of the HS charge transitions Eq. 
(\ref{HS}),
(ii) the optical gap, $\Delta_{\rm opt}$, relevant for the in-gap 
absorption, which opens between the occupied (unoccupied) defect states 
inside the MH gap and the unoccupied (occupied) states of the HS part of 
the UHB (the LHB), and
(iii) the lowest energy excitation gap within the defect states 
$\Delta_{\rm tr}$ Eq. (\ref{tr}), of relevance for transport.
The reported data are obtained for a finite value of the LR Coulomb 
interaction strength $V_{\rm ee}=0.2$ eV, and for two different values 
of the orbital polarization parameter $D$, corresponding to scenarios 
(C) and (D). Note that the gaps $\Delta_{\rm HS}$ and 
$\Delta_{\rm opt}$ in Fig. \ref{fig:25} are inferred from the peak 
positions in the DOS and correspond to the absorption maxima and not 
to the onset of the absorption.

It is evident that only $\Delta_{\rm tr}$ is significantly affected by 
the orbital polarization term (Fig. \ref{fig:24}). We note here that 
while the screening of 
electronic interactions is generally obtained by many-body perturbation 
theory beyond the HF approximation, for example by resummation in random 
phase approximation, the screening of Coulomb interactions at defects is 
already, to a large extent, contained on the Hartree level and taken 
care of by charge and orbital relaxation effects around the defects. 
This explains why finite orbital polarization $D$ leads to such a strong 
reduction of the transport gap.

These characteristic scales may be compared with the experimental data 
\cite{Fuj08} for the doping dependence of the optical absorption spectra
for Y$_{1-x}$Ca$_x$VO$_3$ and La$_{1-x}$Sr$_x$VO$_3$. They reveal HS
Mott transition energies $\Delta_{\rm HS}$ of about $2.2$ eV and $1.8$ 
eV, respectively. The Mott-gap energy is essentially independent of 
doping, as found in the present HF approach. Only close to the MIT, the 
intensity of the MH absorption fades away. The most remarkable feature 
in optical spectroscopy is the appearance of an in-gap absorption 
related to defect states: the intensity of this absorption grows with 
doping. The characteristic energy of the in-gap absorption 
$\Delta_{\rm opt}$ (read off at the peak positions) is independent of 
doping in the dilute limit.

At higher defect concentrations, which go beyond the present model
calculations designed for the dilute limit, i.e., for $x<0.05$, the
energy of the absorption peak decreases and eventually collapses for 
doping concentration approaching the MIT. For example, in the case of 
Y$_{1-x}$Ca$_x$VO$_3$, $\Delta_{\rm opt}\sim 1.2$ eV up to $x=0.10$.
\cite{Fuj08} In La$_{1-x}$Sr$_x$VO$_3$, this absorption is centered at 
about $0.80$ eV for $x<0.10$, while the in-gap absorption peak shifts 
to $\sim 0.3$ eV at the doping concentration $x=0.168$,
i.e., close to the MIT.\cite{Fuj08}
In the high concentration regime it will be important
to include the dielectric screening due to the doped holes.

Activated transport in La$_{1-x}$Sr$_x$VO$_3$ has been reported first by 
Dougier and Hagenmuller,\cite{Dou75} who observed an activation energy 
$\Delta_{\rm tr}\sim 80$ meV at $x\simeq 0.05$ doping. Transport in 
these systems has been extensively discussed by Mott\cite{Mot89} and 
identified as Anderson type, i.e., controlled by defects.
For Y$_{1-x}$Ca$_x$VO$_3$, activated behavior of the resistivity was 
reported for the Ca-doping range $x<0.3$.\cite{Kas93,Sag08,Cin03}
In particular, Sage {\it et al.}\cite{Sag08} reported the activation 
energies $E_a=0.124$, $0.106$, and $0.064$ eV at doping $x=0.1$, 0.2, 
and 0.3, respectively. The values for $\Delta_{\rm tr}$ in Fig. 
\ref{fig:25} are of similar size when orbital polarization $D=0.05$ eV 
is included. Noticing these qualitative trends, we point out here that a
quantitative analysis of the above experimental data will be possible
only after extending the present model to the regime of higher doping
$x>0.1$.

\subsection{Orbital density distribution}
\label{sec:nc}

\begin{figure}[t!]
\includegraphics[width=7.8cm]{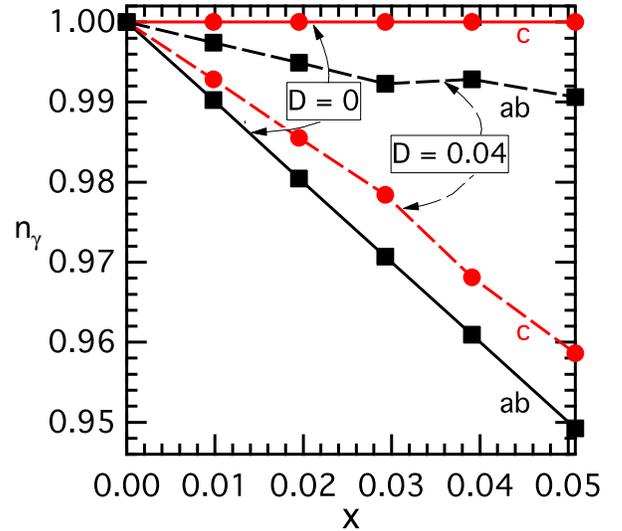}
\caption{(Color online) Average total electron filling $n_\gamma$ in 
the $c$ orbitals ($n_c$, circles) and in the $\{a,b\}$ orbital doublet 
($n_{ab}$, squares), as obtained using an $8 \times 8 \times 8$ cluster 
in the dilute well annealed $CG$ phase, for increasing doping $x$ and 
for two values of the orbital polarization interaction: $D=0$ (solid 
lines), and $D=0.04$ eV (dashed lines). 
Other parameters as in set B of Table I, and $V_{\rm ee}=0$.}
\label{fig:26}
\end{figure}

Spin-orbital order in RVO$_3$ compounds is typically discussed by 
considering only the subspace of $a$ and $b$ orbitals.\cite{Kha01,Ren00}
While such a simplified picture may be sufficient for the undoped 
compounds, it becomes questionable when the system is doped. One 
important and obvious perturbation mixing the orbitals is the 
interaction with the dopands. In fact, this interaction triggers the 
orbital polarization $H_{{\rm pol}}$, induced by the polarization 
constant $D$ in Eq. (\ref{HD}). Figure \ref{fig:26} displays the total 
filling within $c$ and $a+b$ orbitals (defined with respect to the 
original unrotated orbital basis) as function of doping $x$. We see 
that, as expected, the $c$ occupation remains unchanged when the
orbital polarization is absent (at $D=0$), i.e., holes go only into 
the $a$ and $b$ orbitals. Instead, the situation is reversed for 
$D=0.04$ eV and the $c$ occupation is here more strongly reduced than
the $a+b$ occupation.

This, at a first glance, very surprising and counter-intuitive result 
can be understood by recalling that the lowest (occupied) local states 
rotate from pure $c$ character at $D=0$ to a linear combination of $c$ 
and $\{a,b\}$ at finite $D$ in the vicinity of the defects, see Sec.
\ref{sec:atom}. It is important to realize that this rotation involves 
{\it all} V-neighbors of the defect. Thus, the concentration of V-ions 
affected by the redistribution of electronic density is much larger 
than the concentration of doped holes --- this effect explains the 
strong change of occupancy seen in Fig. \ref{fig:26}.

Considering the strong admixture of $a$ and $b$ orbital character in the 
rotated $c'$-orbital, one may wonder whether this does not necessarily 
imply a complete breakdown of any analysis based on a two-flavor 
description. Such concerns are certainly justified when matrix elements 
of local orbitals come into play as, for example, in the calculation of
the intensities of the transitions investigated in optical or RIXS 
spectroscopies. We emphasize that in 
cases where models merely rely on the fact that the rotated $c'$-orbital 
is the lowest orbital in the $t_{2g}$ sector and is always occupied, 
qualitative conclusions based on two-flavor models remain valid.

\begin{figure}[t!]
        \includegraphics[width=8cm]{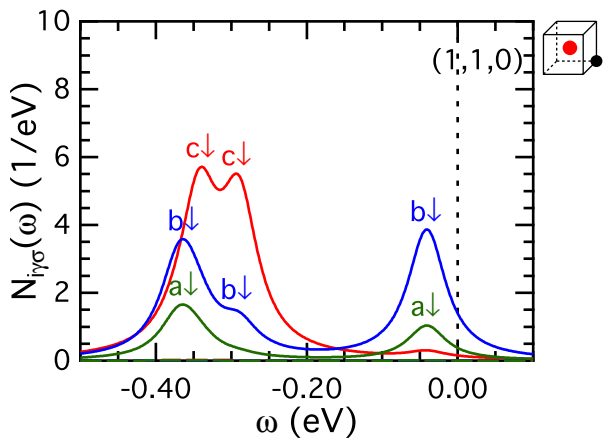}
\caption{(Color online)
Occupied part of the (spin-orbital)-resolved defect state DOSs 
$N_{i\alpha\sigma}(\omega)$ for orbital $\alpha$ and spin $\sigma$ 
at the atom (1,1,0) which belongs to undoped ({\it spectator}) bond, 
as in Fig. \ref{fig:21}(d), obtained in the dilute well annealed doped 
$CG$ phase ($x= 1/64$) using an $8\times 8\times 8$ cluster.
Parameters as in set B of Table \ref{tab:para}, $D=0.05$ eV and 
$V_{\rm ee}=0$.
}
\label{fig:27}
\end{figure}

Further insight into the variation of electronic filling in different
orbitals at increasing doping $x$ can be gained by inspecting the 
occupied single particle states at a spectator site (1,1,0) in the 
vicinity of a defect at $(\frac12,\frac12,\frac12)$ for finite orbital 
polarization $D=0.05$ eV, as displayed in Fig. \ref{fig:27}.
This figure highlights the formation of bonding and antibonding states 
on spectator bonds, which involve the $(a\downarrow)$ and 
$(b\downarrow)$ orbitals as well as the $(c\downarrow)$ orbital 
(but with a much reduced splitting). An important feature here is that 
the chemical potential is 
pinned at the upper edge of the antibonding band. This guarantees that 
the spectator states are blocked for doped holes!

Another important effect of $D$ is that, with respect to the rotated 
basis, the hopping matrix no longer conserves flavors, which is the case  
for the original $t_{2g}$ basis. Therefore, large polarization $D$ 
leads to a strong mixing of orbitals with different character in the 
vicinity of defects also via the kinetic energy. One may wonder which 
mechanism leads to the bonding-antibonding splitting on the spectator 
bond although both states are occupied such that there is no evident 
energy gain as, for instance, in the case of the JT splitting or the 
Peierls distortion.
Here, we observe that the splitting is induced by the orbital rotation
(at finite $D$) and the appearance of the off-diagonal hopping processes 
along the (vertical) $c$ axis. Remarkably, even for the rotated 
$c$ orbital, hopping along the $c$ axis is now possible, as one can see 
from a small splitting of the DOS for $c$ electrons in Fig. \ref{fig:27}.

\section{Summary and conclusions}
\label{sec:summa}

Before summarizing, we recall the fundamental problem in strongly 
correlated systems:
the typical multiplet splitting of the transition metal ions can be 
obtained in the localized limit by a Hartree factorization performed 
with respect to an optimal local basis set, i.e., a basis with 
occupation numbers being close to either 1 or 0. That is, the local
spin-orbital states should be either occupied or empty --- intermediate 
values for the occupation numbers would signal the breakdown of the Mott 
gap in the Hartree approximation, which ultimately happens when the 
kinetic energy increases and dominates over electron localization 
triggered by the local interactions. The former requirement, however, 
leads to complications in the vicinity of defects. This follows from  
a rotation of the occupied orbital states of the V-ions due to the 
local orbital-polarization fields associated with charged defects.

In principle, a local rotation could be specified that removes the 
off-diagonal polarization terms and defines a new optimal local basis,
i.e., again with occupancies either close to 1 or 0, and a fully
developed Mott gap. This simple scheme is not applicable when some 
nonlocal terms are present (as e.g. the kinetic energy term). Yet, 
it is important to recognize that the unrestricted Hartree-Fock (HF) 
(and not Hartree) scheme, i.e., including the relevant off-diagonal 
contributions from (local and non-local) interactions, can be used to 
determine the Mott-Hubbard split bands in the presence of defects. We 
have found that this also holds true for the three-flavor case. In 
particular, we have shown, by comparison with exact diagonalization, 
that the HF approach provides a surprisingly faithful description not 
only of the occupied, but also of the unoccupied higher multiplet states. 
It is the latter aspect that is particularly surprising here as it is 
known that the HF approach is designed to describe the occupied states 
in the best possible way, while for the unoccupied states one expects a 
considerably poorer description (e.g. electron affinities).

Basic features of the character of wave functions and excitation 
spectra, which reflect the interplay of strong correlations and the 
orbital polarization field $D$ due to a defect, are already well 
represented by the HF study of a single ion or a bond 
embedded into the electronic structure of a large cluster reflecting 
the Mott insulator with spin-orbital order. These studies 
and their comparison with exact diagonalization allow us to conclude 
that the HF calculation describes well: 
(i) the effect of the orbital polarization on the wave functions in 
the three-flavor case, and 
(ii) the multiplet structure for an atom (Sec. \ref{sec:atom}). 
A further investigation for an embedded bond (Sec. \ref{sec:bond}) 
highlights the interplay of the orbital polarization field $D$ and the 
kinetic energy. Comparison of the HF results for partial densities and 
excitation spectra (i.e., photoemission and inverse photoemission) with 
exact diagonalization data shows good agreement in the whole $D$ range. 
An exception are extremely small values of $D$, where the exact wave 
function is entangled, a feature that 
cannot be captured within the HF. A particularly good agreement between 
the HF and exact diagonalization is obtained for the binding energy 
basically for all values of $U>2$ eV. The binding energy for the $CG$ 
phase is found to be much larger than that for the $GC$ phase.
This higher binding energy, together with the double exchange mechanism
realized on ferromagnetic bonds, is responsible for the increasing
stability of the $CG$ phase on increasing the doping and leads to the
experimentally observed magnetic phase transition in
Y$_{1-x}$Ca$_x$VO$_3$ at very low doping, $x\sim 0.02$.\cite{Hor11}

The description of doped carriers in Mott insulators, such as in the 
$R_{1-x}A_x$VO$_3$ compounds ($R$=Y, La and $A$=Ca, Sr, etc.) or in 
LaVO$_3$/SrVO$_3$ superlattices,\cite{Lud09} is a computational challenge 
due to the interplay of strong correlations, defect-induced local 
deformations of the wave functions and spin-orbital order. It is evident 
that to deal with strong correlations and defects, for sufficiently large 
systems, requires carefully chosen approximations. In this work, we have 
shown that the unrestricted HF method fulfils all essential requirements. 
In particular, we have seen that even the unoccupied states at high 
energy, i.e., those reflecting the multiplet structure of the Mott 
insulator at orbital degeneracy, are well described in a doped system. 
The same holds true for the perturbed electronic structure in the 
vicinity of the charged defects. The HF approach is an efficient scheme 
which maps the interacting electron problem onto the problem of a single 
particle moving in a self-consistently determined field of the other 
electrons. Therefore, it is applicable for large systems, systems with 
defects, heterostructures and interfaces. Very relevant for the success 
of the method in describing Mott insulators are the broken symmetries
reflecting the underlying spin-orbital order. Fortunately, the 
spin-orbital order in $R_{1-x}A_x$VO$_3$ compounds persists up to the 
metal-insulator transition, thus the assumption of spin-orbital order 
in the unrestricted HF scheme is here fully justified.

Having shown that the HF approach provides a faithful description
of excited states of single defects in systems with broken symetries, 
we investigated the electronic structure of an $8\times8\times8$ 
cluster for finite defect concentrations in the dilute doping regime. 
The aim here is to resolve the complex nature of defect states and 
their manifestation in the excitation spectra. A single $A$ defect, 
for example a divalent $A$ ion in $R_{1-x}A_x$VO$_3$, adds a hole 
into the $t_{2g}$ shell of V($3d$) states. In the dilute limit, this 
hole is not freely moving but is pinned by the charge of the defect 
essentially to a cube of V ions surrounding the defect $A$ in its 
center. 

The present study demonstrates that the excitation spectra of 
the intrinsic, orbital-degenerate Mott insulator are strongly modified 
by adding defects. We address here only some points relevant for  
experimental studies. An important observation is that the fundamental 
excitations, such as the Mott-gap and the multiplet energies, are not 
affected by doping in the dilute regime. The defects generate defect 
states inside the Mott-Hubbard gap, as observed.\cite{Fuj08} Due to the 
intrinsic doping, the chemical potential lies inside the defect band.
The spectral weight of defect states is taken from the LHB.
We have argued that the in-gap absorption observed in the optical 
conductivity experiments performed for several doped vanadate systems 
can be identified with these states.

Moreover, we observe in the HF excitation spectra for the $CG$ phase
a transport gap inside the defect band, which corresponds to the removal 
of an electron at one defect and the addition at another defect in the 
neighborhood. In the absence of electron-electron interactions, this gap 
is essentially determined by a bonding-antibonding splitting resulting 
from the $c$-axis kinetic energy of a doped hole confined by the defect.
The inclusion of LR Coulomb interactions between electrons 
$\sim V_{\rm ee}$ leads to an approximately linear increase of the 
transport gap with $V_{\rm ee}$.
Thus, these interactions promote the transport gap to a Coulomb gap.
Combining Coulomb interactions and the mechanism of orbital polarization 
we find a reduction of the size of the Coulomb gap. This can be 
considered as screening in the vicinity of the localized defect, i.e., 
screening contained already at the HF level.

An important motivation of this work was the study of the three-flavor 
case, particularly in connection with defects in the $R$VO$_3$ 
perovskites. Usually spin-orbital order in the perovskite vanadates is 
discussed in terms of a simplified two-flavor ($\{a,b\}$) model.
\cite{Hor11} We have shown here that defects lead indeed to a strong 
change of occupied orbitals, e.g. the occupied $c$-orbitals, in the 
vicinity of charged defects due to orbital 
polarization. Yet, we have also shown that the topmost occupied local 
orbitals, i.e., the orbitals that are relevant for the doped holes, are 
mainly of $\{a,b\}$-character as in the two-flavor description. 
It is this latter observation that leads to the conclusion that the 
interaction of doped holes and the spin-orbital degrees of freedom are 
in fact similar in the two models.

Summarizing, the central result of this paper is establishing that the 
unrestricted HF approach is well designed to describe the electronic 
structure of charged defects in doped transition metal oxides with 
active orbital degrees of freedom, in presence of strong electron 
correlations. The present study provides valuable insights into the 
changes of the electronic structure in doped Mott-Hubbard systems 
under doping. While the positions of the Hubbard subbands are not 
affected by doping, their spectral intensity changes and new defect
states occur within the Mott-Hubbard gap. These new states are observed
in the optical spectroscopy and their weight increases with doping.

Finally, we remark that the HF approach represents a natural basis for 
a subsequent many-body perturbative treatment. It should be noted that 
the unrestricted HF method already provides the Mott-Hubbard gap and 
the higher energy multiplet excitations. Thus, the proper many-body 
treatment would start from a reasonable Slater determinant, which 
reflects the correlated electronic structure of the spin-orbital ordered 
Mott insulator. We have shown that the fundamental Mott gap is correctly 
described by the HF in the framework of the multiband Hubbard model. 
Therefore, the many-body treatment will hardly affect the Mott-Hubbard 
gap, yet it may further improve higher multiplets, and it will certainly 
contribute to the screening of the long-range Coulomb interactions and 
lead to an extra reduction of the Coulomb gap in the defect states.

\acknowledgments

We thank Raymond Fr\'esard for insightful discussions and
Giniyat Khaliullin for careful reading of the manuscript.
A.A. thanks the Max-Planck-Institut f\"ur Festk\"orperforschung,
Stuttgart for hospitality and financial support.
A.M.O. acknowledges support by the Polish National Science 
Center (NCN) under Project No. 2012/04/A/ST3/00331.

\appendix*

\section{Hartree versus Hartree-Fock approximation and optimal basis}
\label{sec:gen}

The remarkable successes of the LDA+$U$ approach\cite{Ani91} in 
describing the various aspects of the electronic structure of strongly 
correlated materials,\cite{Ani97,Sol08,Ima10} including the magnetic 
order and the MH gaps, suggest that the Hartree (or mean field) 
approximation for the electron-electron interactions in the appropriate 
orbital basis could be sufficient to design an efficient and realistic 
scheme to determine the electronic structure. This experience, however, 
is based on undoped systems, where the occupation numbers for the 
{\it atomic} orbitals (such as $t_{2g}$ or $e_g$ orbitals at transition 
metal ions in the correlated oxides) used in the LDA+$U$ calculations 
are either close to 1 or close to 0. This latter condition is actually 
necessary to produce an essentially correct multiplet splitting, i.e., 
these orbitals are already the properly chosen orbitals to implement the
electron interactions in the Hartree scheme. This is always the case
in an undoped Mott insulator, which explains the success of the LDA+$U$
approach.\cite{Ani97,Sol08,Ima10} Note also that the orbital
basis is then identical at every site.

However, in a system with defects the occupied orbitals (here we 
consider $\downarrow$-spin ones) belong to an orbital basis, which 
adjusts itself due to the external polarization field (\ref{HD}), 
see Sec. \ref{sec:atom}. As a result, for an atom one finds new rotated 
orbitals $\{|\xi_n\rangle\}$, introduced in Sec. \ref{sec:atom}. 
Similar situation occurs in a bond but in addition the orbitals are then 
delocalized over two atoms in the $CG$ phase, see Sec. \ref{sec:bond}. 
Hence we write the HF orbitals for $\downarrow$-spin electrons 
$\{|\xi_n\rangle\}$ as a linear combination of the original $t_{2g}$ 
orbitals, $\{|ia\rangle,|ib\rangle,|ic\rangle\}$ (\ref{abc}) at site $i$,
\begin{equation}
|\xi_n\rangle=
\sum_{i}\sum_{\gamma=a}^c\alpha_{i\gamma}^{(n)}|i\gamma\rangle\,.
\label{c'}
\end{equation}
Here $i=1$ and $n=1,3,5$ for an atom, while $i=1,2$ and $n=1,\dots,6$ 
for a bond along the $c$ axis. As an example, we give in Table
\ref{tab:exp} the expansion coefficients,
\begin{equation}
|\alpha_{i\gamma}^{(n)}|^2\equiv |\langle\xi_n|i\gamma\rangle|^2\,
\label{exp}
\end{equation}
obtained for a representative value of $D=0.05$ eV.
These coefficients were obtained by applying the unrestricted HF
approximation as described in Secs. \ref{sec:atom} and \ref{sec:bond}.

It is important to realize that the Fock terms that stem from
interorbital Coulomb interactions between electrons with the same spin
play an essential role in optimizing the orbital basis near charge
defects. Indeed, the polarization term (\ref{HD}) acts as a field and
contains on-site interorbital averages,
$\langle c_{i\alpha\sigma}^\dagger c_{i\beta\sigma}^{}\rangle$ with 
$\alpha\neq\beta$, which are responsible for the orbital rotation and 
the change of the local basis. By comparing the occupied orbitals for 
an atom and for a bond in the $C$-AF phase at finite value of $t=0.2$ 
eV, see Table \ref{tab:exp}, one finds that further change of the 
orthogonal basis occurs at each site due to finite kinetic energy along 
the FM bond.\cite{noteaf} Note that the orbital labels $\{n\}$ in Table 
\ref{tab:exp} are assigned to the orbital functions $\{\xi_n\}$ in 
sequence of their increasing HF energy. The orbitals $|\xi_1\rangle$ 
and $|\xi_3\rangle$ are occupied for an atom, while for the bond the 
number of occupied orbitals is doubled and also the orbitals 
$|\xi_2\rangle$ and $|\xi_4\rangle$ are occupied --- they are 
characterized by similar orbital densities as those in the orbitals 
$|\xi_1\rangle$ and $|\xi_3\rangle$ for the atom, respectively, but the 
energies of the latter two orbitals are interchanged in the bond.

\begin{table}[t!]
\caption{
Decomposition of the local orbital basis Eq. (\ref{c'}) onto the original 
$t_{2g}$ basis given by the elements $|\alpha_{i\gamma}^{(n)}|^2$ 
(\ref{exp}) for $\downarrow$-spin electrons,
as obtained for an atom (atom, $n=1,3,5$) and for a bond with two atoms
$i=1,2$ (bond, $n=1,\cdots,6$) in the $CG$ phase, with $D=0.05$ eV.
Other parameters as in Table I: set A (atom) and set B (bond).}
\begin{ruledtabular}
\begin{tabular}{cccccccc}
&         &          &  atom  &  \multicolumn{3}{c}{bond} &\cr
& orbital & $\gamma$ &        &  $i=1$ &  $i=2$ &  total &\cr
\colrule
& $\xi_1$ &   $c$    & 0.8616 & 0.0058 & 0.0058 & 0.0116 &\cr
&         &   $b$    & 0.0832 & 0.3100 & 0.1842 & 0.4942 &\cr
&         &   $a$    & 0.0552 & 0.1842 & 0.3100 & 0.4942 &\cr
& $\xi_2$ &   $c$    &   ---  & 0.4117 & 0.4117 & 0.8234 &\cr
&         &   $b$    &   ---  & 0.0556 & 0.0326 & 0.0882 &\cr
&         &   $a$    &   ---  & 0.0326 & 0.0556 & 0.0882 &\cr
& $\xi_3$ &   $c$    & 0.0032 & 0.4337 & 0.4337 & 0.8674 &\cr
&         &   $b$    & 0.5381 & 0.0430 & 0.0232 & 0.0662 &\cr
&         &   $a$    & 0.4587 & 0.0232 & 0.0430 & 0.0662 &\cr
& $\xi_4$ &   $c$    &   ---  & 0.0096 & 0.0096 & 0.0192 &\cr
&         &   $b$    &   ---  & 0.3293 & 0.1611 & 0.4904 &\cr
&         &   $a$    &   ---  & 0.1611 & 0.3293 & 0.4904 &\cr
& $\xi_5$ &   $c$    & 0.1352 & 0.0786 & 0.0786 & 0.1572 &\cr
&         &   $b$    & 0.3787 & 0.1150 & 0.3063 & 0.4213 &\cr
&         &   $a$    & 0.4861 & 0.3063 & 0.1150 & 0.4213 &\cr
& $\xi_6$ &   $c$    &   ---  & 0.0605 & 0.0605 & 0.1210 &\cr
&         &   $b$    &   ---  & 0.1470 & 0.2925 & 0.4395 &\cr
&         &   $a$    &   ---  & 0.2925 & 0.1470 & 0.4395 &\cr
\end{tabular}
\end{ruledtabular}
\label{tab:exp}
\end{table}

The local electron-electron interactions are rotationally invariant,
\cite{Ole83} and the Hartree scheme could in principle be applied to
any set of locally orthogonal orbitals including the basis of rotated
orbitals discussed above. If one could find such a rotated basis with 
the self-consistently determined orbital occupations being just either 
0 or 1, Hartree approximation would again be again sufficient as in the 
LDA+$U$ scheme. Mean fields would then suffice to come quite close to 
the exact solution (as for a single atom, see Sec. \ref{sec:atom}), and 
to reproduce the multiplet structure of the UHB.
In such a case, Fock terms simply vanish and one can neglect them from
the very beginning despite the fact that terms that could drive finite
values of the off-diagonal averages are present in the HF Hamiltonian. 

Summarizing, we conclude that the off-diagonal Fock elements are 
essential within the HF scheme and are necessary to arrive at the 
final optimal orbitals adjusted to the defect states.
It is only in this orbital basis that the electronic structure in a Mott 
insulator faithfully reproduces the multiplet splittings. Therefore, 
possible extensions of the LDA+$U$ scheme to doped Mott insulators will 
have to use downfolding procedure\cite{down} to a proper tight-binding 
model\cite{Gun12} in which full unrestricted HF calculations could be 
performed. Designing such a scheme which has to go {\it beyond} the 
present LDA+$U$ approach\cite{Ani91} is indeed very challenging --- 
it would make it possible to investigate
not only doped Mott insulators but also interfaces,\cite{Hwa12,Pav12}
heterostructures,\cite{Lud09} or other composite materials with defect 
states in correlated insulators in the future.


\begin{thebibliography}{199}
\bibitem{notecg} We call the phase with $G$-type antiferromagnetic (N\'eel)
                 order accompanied by $C$-type orbital order (with the same
                 occupied orbitals along the $c$ cubic axis and alternating 
                 orbitals within the $ab$ plane) a $GC$ phase, and the phase 
                 with $C$-type antiferromagnetic order (ferromagnetic spin  
                 order along the $c$ cubic axis and staggered moments within 
                 the $ab$ plane) and staggered orbitals in all three directions 
                 --- a $CG$ phase.
\bibitem{notew} Using the parameter set B of Table I, one obtains energies 
   per site: $E_{CG}=-2.241$ eV and $E_{GC}=-2.200$ eV. The $CG$ phase has 
   then lower total energy and this situation is realized in LaVO$_3$, 
   where the $CG$ phase is more stable. In contrast, the available 
   experiments have concluded that the $GC$ phase is more stable in YVO$_3$. 
   This could be easily obtained within the present model by a proper fine 
   tuning of the model parameters. At any rate, on slightly increasing the
   doping, $x\sim 0.02$, the $GC$ phase destabilizes in Y$_{1-x}$Ca$_x$VO$_3$
   too in favor of the $CG$ phase. This phase transition was qualitatively
   discussed in the two-flavor model,\cite{Hor11} where it has been shown
   that increasing stability of the $CG$ phase with doping is generic and 
   follows from the double exchange mechanism.\cite{deG60,vdB99,Hor99,Ber12}
\bibitem{notex} These excitations are obtained by diagonalizing the
                on-site Hamiltonian Eq. (\ref{HU}) in the subspace of
                three degenerate $t_{2g}$ orbitals filled by 
                $m=2$ electrons.
\bibitem{noteu} This choice avoids the unphysical regime of $U<1.8$ eV where 
                the first excitation energy Eq. (\ref{hs}) is negative.
\bibitem{noteaf} For AF bonds the hopping is blocked by double exchange 
                 \cite{deG60,vdB99,Hor99,Ber12} so this orbital modification  
                 by the hopping processes is absent as long and the magnetic 
                 order persists, at least locally.
\end{thebibliography}
\end{document}